\documentclass{tlp}
\usepackage{latexsym}
\usepackage{harvard}
\usepackage{epsf}

\def\mymath#1{\relax\ifmmode#1\else$#1$\fi}

\newcommand{\seq}[1]{\mymath{\overline{#1}}}
\newcommand{\svbt}[1]{{\tt\small #1}}
\newcommand{\mikearraystretch}{1}

\newcommand{\pdset}[1]{{\cal #1}}

\newcommand{\specialise}{\leadsto}

\newtheorem{examplerm}{Example}
\newenvironment{example}{\begin{examplerm}\rm}{\end{examplerm}}

\newtheorem{definitionrm}{Definition}
\newenvironment{definition}{\begin{definitionrm}\rm}{\end{definitionrm}}

\newtheorem{propositionrm}{Proposition}
\newenvironment{proposition}{\begin{propositionrm}\rm}{\end{propositionrm}}

\newtheorem{algorithmrm}{Algorithm}
\newtheorem{procedurerm}{Procedure}

\newenvironment{procedure}{\begin{procedurerm}\rm}{\end{procedurerm}}

\newcommand{\larr}{\leftarrow}

\def\mymath#1{\relax\ifmmode#1\else$#1$\fi}

     {\begin{list}{--}{
     \setlength{\itemsep}{0 pt}
     \setlength{\parsep}{0 pt}
     \setlength{\topsep} {0 pt} }}
     {\end{list}}
\newenvironment{pitemize}
     {\begin{list}{}{
     \setlength{\itemsep}{0 pt}
     \setlength{\parsep}{0 pt}
     \setlength{\topsep} {2 pt} }}
     {\end{list}}

\newcommand{\ignore}[1]{} 
\newcommand{\forExperts}[1]{} 

\newcommand{\condtext}[2]{#1}

\newcommand{\system}[1]{{\sc #1}}

\newenvironment{SProg}
     {\begin{small}\begin{tt}\begin{tabular}[c]{l}}%
     {\end{tabular}\end{tt}\end{small}}
\newcommand{\qin}{\hspace*{0.15in}}

\begin{document}

\bibliographystyle{dcu}

\title[Offline Specialisation in Prolog]{Offline Specialisation in Prolog Using a
 Hand-Written Compiler Generator}
\author[M.\ Leuschel et al.]
{MICHAEL LEUSCHEL\\
DSSE, Department of Computer Science,
University of Southampton\\ Highfield, SO17 1BJ, UK
\and
 JESPER J{\O}RGENSEN\\
Dept.~of Mathematics and Physics,
Royal Veterinary and
Agricultural University,\\
Thorvaldsensvej 40, DK-1871 Frederiksberg C, Denmark
\and
 WIM VANHOOF ~ and ~  MAURICE BRUYNOOGHE\\
Department of Computer Science, Katholieke Universiteit Leuven\\
Celestijnenlaan 200A, B-3001 Heverlee, Belgium}
\pagerange{\pageref{firstpage}--\pageref{lastpage}}
\volume{\textbf{10} (3):}
\jdate{?? 200?}
\setcounter{page}{1}
\pubyear{200?}

\maketitle
\label{firstpage}

\begin{abstract}
  The so called ``$\mathit{cogen}$ approach'' to program
  specialisation, writing a compiler generator instead of a
  specialiser, has been used with considerable success in partial
  evaluation of both functional and imperative languages.  This paper
  demonstrates that the $\mathit{cogen}$ approach is also applicable
  to the specialisation of logic programs (called partial deduction)
   and leads to effective
  specialisers. Moreover, using good binding-time annotations, 
  the speed-ups of the specialised programs are
  comparable to the speed-ups
  obtained with online specialisers.
  
  The paper first develops a generic approach to offline partial
  deduction and then a specific offline partial deduction method,
  leading to the offline system {\sc lix} for pure logic programs.
  While this is a usable specialiser by itself, it
   is used to develop the $\mathit{cogen}$ system {\sc logen}.
  Given a program, a specification of what inputs will be static, and
  an annotation specifying which calls should be unfolded, {\sc logen} 
  generates a specialised specialiser for the program at hand. Running
  this specialiser with particular values for the static inputs
  results in the specialised program. While this requires two steps
  instead of one, the efficiency of the specialisation process is
  improved in situations where the same program is specialised
  multiple times.
  
  The paper also presents and evaluates an automatic binding-time
  analysis
  that is able to derive the annotations. While the derived  
  annotations are still suboptimal compared to hand-crafted ones, 
  they enable non-expert users to use the {\sc logen} system
  in a fully automated way.
  
  Finally, {\sc logen} is extended so as to directly support a large
  part of Prolog's declarative and non-declarative features and so as
  to be able to perform so called mixline specialisations.

{\bf Keywords} Partial evaluation, partial deduction, program specialisation,
 compiler generation, abstract interpretation
\end{abstract}



\section{Introduction}

{\em Partial evaluation\/} has over the past decade received
considerable attention both in functional (e.g.\ \cite{Jones:peval}),
imperative (e.g.\ \cite{Andersen94:PhD}) and logic programming (e.g.\ 
\cite{Gallagher:PEPM93,komorowski:meta92,Pettorossi94:jlp}).  Partial
evaluators are also sometimes called $\mathit{mix}$, as they usually
perform a mixture of evaluation and code generation steps.  In the
context of pure logic programs, partial evaluation is sometimes
referred to as {\em partial deduction\/}, the term partial evaluation
being reserved for the treatment of impure logic programs.

Guided by the {\em Futamura projections\/} \cite{Futamura71} a lot of
effort, specially in the functional partial evaluation community, has
been put into making systems self-applicable. A partial evaluation or
deduction system is called {\em self-applicable\/} if it is able to
effectively%
\footnote{This implies some efficiency considerations, e.g.\ the
  system has to terminate within reasonable time constraints, using an
  appropriate amount of memory.}%
specialise itself.  In that case one may, according to the second
Futamura projection, obtain {\em compilers} from interpreters and,
according to the third Futamura projection, a {\em compiler generator}
($\mathit{cogen}$ for short).  In essence, given a particular program
$P$, a $\mathit{cogen}$ generates a {\em specialised specialiser\/}
for $P$.  If $P$ is an interpreter a $\mathit{cogen}$ thus generates a
compiler.

However writing an effectively self-applicable specialiser is a
non-trivial task --- the more features one uses in writing the
specialiser the more complex the specialisation process becomes,
because the specialiser then has to handle these features as well.
This is why so far no partial evaluator for full Prolog (like
\system{mixtus}\ \cite{Sahlin93:ngc}, or \system{paddy}\ 
\cite{Prestwich92:TR}) is effectively self-applicable. On the other
hand a partial deducer which specialises only purely declarative logic
programs (like \system{sage} \cite{Gurr:PHD} or the system in
\cite{Bondorfetal:90}) has itself to be written purely declaratively
leading to slow systems and impractical compilers and compiler
generators.

So far the only practical compilers and compiler generators for logic
programs have been obtained by \cite{FujitaFurukawa88:ngc} and
\cite{MogensenBondorf:LOPSTR92}. However, the specialisation in
\cite{FujitaFurukawa88:ngc} is incorrect with respect to some
extra-logical built-ins, leading to incorrect results when attempting
self-application \cite{Bondorfetal:90}.  The partial evaluator
\system{logimix}\ \cite{MogensenBondorf:LOPSTR92} does not share this
problem, but gives only modest speedups when self-applied (compared to
results for functional programming languages; see
\cite{MogensenBondorf:LOPSTR92}) and cannot handle partially static
data.

However, the actual creation of the $\mathit{cogen}$ according to the
third Futamura projection is not of much interest to users since
$\mathit{cogen}$ can be generated once and for all when a specialiser
is given. Therefore, from a user's point of view, whether a
$\mathit{cogen}$ is produced by self-application or not is of little
importance; what is important is that it exists and that it is
efficient and produces efficient, non-trivial specialised
specialisers. This is the background behind the approach to program
specialisation called the {\em cogen approach} (as opposed to the more
traditional $\mathit{mix}$ approach): instead of trying to write a
partial evaluation system $\mathit{mix}$ which is neither too
inefficient nor too difficult to self-apply one simply writes a
compiler generator directly.  This is not as difficult as one might
imagine at first sight: basically the $\mathit{cogen}$ turns out to be
just a simple extension of a ``binding-time analysis'' for logic
programs (something first discovered for functional languages in
\cite{Holst:89a} and then exploited in, e.g.,
 \cite{HolstLaunchbury:92,%
BirkedalWelinder:94:handcogen,Andersen94:PhD,GlueckJoergensen:PLILP95,%
Thiemann:ICFP96}).

In this paper we will describe the first $\mathit{cogen}$ written in
this way for a logic programming language.  We start out with a
$\mathit{cogen}$ for a small subset of Prolog and progressively
improve it to handle a large part of Prolog and to extend its
capabilities.

Although the Futamura projections focus on how to generate a compiler
from an interpreter, the projections of course also apply when we
replace the interpreter by some other program.
In this case the program produced by the second Futamura projection is
not called a compiler, but a {\em generating extension}. The program
produced by the third Futamura projection could rightly be called a
{\em generating extension generator} or gengen, but we will stick to
the more conventional $\mathit{cogen}$.

The main contributions of this work are:
\vspace{1ex}

\begin{itemize}
  
\item[1.] A formal specification of the concept of {\em binding-time
    analysis\/} and more generally {\em binding-type analysis\/},
  allowing the treatment of {\em partially static\/} structures, in a
  (pure) logic programming setting and a description of how to obtain
  a generic procedure for {\em offline partial deduction\/} from such
  an analysis.
  
\item[2.] Based upon point 1, the first description of an efficient,
  handwritten compiler generator ($\mathit{cogen}$) for a logic
  programming language, which has --- exactly as for other handwritten
  cogens for other programming paradigms --- a quite elegant and
  natural structure.

\item[3.] A way to handle both {\em extra-logical\/} features (such as
  {\tt\small var/1}) and {\em side-effects\/} (such as {\tt\small
    print/1}) within the $\mathit{cogen}$. A refined treatment of the
  {\tt\small call/1} predicate is also presented.
  
\item[4.] How to handle negation, disjunction and the if-then-else
  conditional in the $\mathit{cogen}$.
  
\item[5.] Experimental results showing the efficiency of the
  $\mathit{cogen}$, the generating extensions, and also of the
  specialised programs.

\item[6.] A method to obtain a binding-type analysis through
 the exploitation of existing termination analysers.
\end{itemize}
\vspace{1ex}

This paper is a much extended and revised version of
\cite{JorgensenLeuschel:Cogen}: points 3, 4, 5, 6 and the partially
static structures of point 1 are new, leading to a more powerful and
practically useful $\mathit{cogen}$.

The paper is organised as follows: 
In Section~\ref{section:off-line-pd} we formalise the concept of
 off-line partial deduction and the associated binding-type analysis.
In Section~\ref{section:cogen-pure} we present and explain our
 $\mathit{cogen}$ approach in a pure logic programming setting,
  starting from the structure of the generating extensions.
In Section~\ref{section:cogen-impure}
 we discuss the treatment of declarative and non-declarative 
built-ins as well as
 constructs such as negations, conditionals,
and  disjunctions.
In Section~\ref{section:results} we present experimental results
 underlining the efficiency of the $\mathit{cogen}$ and of the 
 generating extensions it produces.
We also compare the results against a traditional offline specialiser.
In Section~\ref{section:bta}
 we present a method for doing an automatic binding-type
 analysis.
We evaluate the efficiency and quality of this approach using
 some experiments.
We conclude with some discussions of related and future work
 in Section~\ref{section:discussion}.


\section{Off-line Partial Deduction}
\label{section:off-line-pd}


Throughout this paper, we suppose familiarity with basic notions in logic
 programming.
We follow the
 notational conventions of \cite{Lloyd:flp}.
In particular, in programs, we denote variables through
strings starting
 with an upper-case symbol, while the notations
 of constants, functions and predicates begin with a lower-case character.

\ignore{****
We will also use the following not so common notations.
Given a function $f: A\mapsto B$
 we often use the
 {\em natural extension\/} of $f$,
 $f^{*}: 2^{A}\mapsto 2^{B}$, defined by 
  $f^{*}(S) = \{f(s) \mid s\in S\}$%
\footnote{Like {\tt map} in functional languages.}.
Similarly, given a function $f: A\mapsto 2^{B}$ we also
 define the function
 $f_{\cup}: 2^{A}\mapsto 2^{B}$, by $f_{\cup}(S) = \cup_{s\in S}f(s)$.
Given a function $f: A\times B \mapsto C$ and an element $a\in A$
 we define the curried version of $f$, $f_{a}: B\mapsto C$, by
 $f_{a}(X) = f(a,X)$.
Finally, we will denote by $A\rightarrow A ;  A$
the Prolog conditional.
*****}


\subsection{A Generic Partial Deduction Method}

We start off by presenting a general procedure for performing partial deduction.
More details on 
 partial deduction and how to control it
 can be found, e.g., in \cite{LeuschelBruynooghe:TPLP02}.

Given a logic program $P$ and
 a goal $G$, {\em partial deduction\/} produces a new program $P'$ which is
 $P$ ``specialised'' to the goal $G$;
 the aim being that the specialised program $P'$ is more efficient than
 the original program $P$ for all goals which are instances of $G$.
The underlying technique of partial deduction
 is to construct finite, non-trivial but possibly incomplete
 SLDNF-trees.
(A {\em trivial\/} SLDNF-tree is one in which no literal in
  the root has been selected for resolution,
   while an {\em incomplete\/} SLDNF-tree is a SLDNF-tree which,
 in addition to success
 and failure leaves, may also contain leaves where no literal has
 been selected for a further derivation step.)
The derivation steps in these SLDNF-trees
 correspond to the computation steps which
 have already been performed by the 
 {\em partial deducer\/}\index{partial deducer} and
the clauses of the specialised program are then extracted from these trees
 by constructing one specialised clause
 (called a {\em resultant}) per non-failing branch.
These SLDNF-trees and resultants are obtained as follows.

\begin{definition}
\label{def:unfoldingrule}
An {\em unfolding rule\/} is a function which,
 given a program $P$ and a goal $G$,
 returns a
 non-trivial and possibly incomplete
 SLDNF-tree for $P\cup\{G\}$.
\end{definition}

\begin{definition} 
\label{def:resultants}
Let $P$ be a normal program and $A$ an atom. Let $\tau$ be a finite,
  incomplete
 SLDNF-tree for $P \cup \{\leftarrow A\}$.
Let
 $\leftarrow G_{1},\ldots,\leftarrow G_{n}$ be the goals in the
 leaves of the non-failing branches of $\tau$.
Let $\theta_{1},\ldots,\theta_{n}$ be the computed
answers of the derivations from $\leftarrow A$ to
$\leftarrow G_{1},\ldots,\leftarrow G_{n}$ respectively.
Then the set of resultants, $resultants(\tau)$, is defined to
 be the set of clauses
 $\{A\theta_{1}\leftarrow G_{1},\ldots, A\theta_{n}\leftarrow G_{n}\}$.
We also define the set of leaves, $\mathit{leaves}(\tau)$, to be
 the atoms occurring in the goals $G_{1},\ldots,G_{n}$.
\end{definition}

Partial deduction 
 uses the resultants for a given set of atoms $\pdset{S}$ to
 construct the specialised program (and for each atom in $\pdset{S}$
 a different specialised predicate definition will be generated).
Under the conditions stated in \cite{Lloyd:jlp91},
namely {\em closedness\/} (all leaves are an instance of an
 atom in $\pdset{S}$) and 
 {\em independence} (no two atoms in $\pdset{S}$ have a common instance),
 correctness of the specialised program is guaranteed.

In most practical approaches
 independence is ensured by using
 a {\em renaming\/} transformation which
 maps dependent atoms to new predicate symbols.
Adapted correctness results can be found in \cite{Benkerimi93:jlc,%
LeuschelMartensDeSchreye:Toplas}
 and \cite{LeuschelDeSchreyeDeWaal:JICSLP96}.
Renaming is often combined with argument {\em filtering\/} to improve the
 efficiency of the specialised program;
 see e.g.\ \cite{Gallagher:META90,Benkerimi93:jlc,%
LeuschelSorensen:RAF}.

Closedness can be ensured by using the following
 outline of a partial deduction procedure,
 similar to the ones used in e.g.\ 
\cite{Gallagher91:TR,Gallagher:PEPM93,%
LeuschelDeSchreye97:ngc}.%

\begin{procedurerm}{\vspace*{-0.48cm}~~~~~~~~~~~~~~~~~~(Partial deduction)}
\label{skeleton:pdalgorithm}
\ignore{
\begin{enumerate}
\item Let $\pdset{S}_{0}$ be the set of atoms to be specialised and let
 $i=0$.
\item \label{step:pdalgo-unfold}
 Apply the unfolding rule $U$ to each element of $\pdset{S}_{i}$:
        $\Gamma_{i} = U_{P}^{*}(\pdset{S}_{i})$.
\item $S_{i+1} = \mathit{generalise}(\pdset{S}_{i}\cup \mathit{leaves}_{\cup}(\Gamma_{i}))$
\item
If $\pdset{S}_{i+1} \neq \pdset{S}_{i}$ (modulo variable renaming)
 increment $i$ and restart
 at step~\ref{step:pdalgo-unfold},
 otherwise generate the specialised program by
 applying a renaming (and filtering)
 transformation to $resultants_{\cup}(\Gamma_{i})$.
\end{enumerate}
}
\begin{description}
\item[{\bf Input}:] a program $P$ and an initial set ${\cal S}_{0}$ of atoms
 to be specialised
\item[{\bf Output}:] a set of atoms $\pdset{S}$
\item[{\bf Initialisation}:] ${\pdset{S}}_{new}:=\mathit{generalise}({\cal S}_{0})$
\item[{\bf repeat}] \mbox{ }
\vspace*{-1ex}
  \begin{description}
  \item[ ] $\pdset{S}_{old}:=\pdset{S}_{new}$
  \item[ ] $S_{new} := \{ s_{n} \mid s_{n}\in \mathit{leaves}(\mathit{unfold}(P,s_{o})) \wedge$
      $s_{o}\in\pdset{S}_{old} \}$
  \item[ ] $S_{new} := \mathit{generalise}(\pdset{S}_{old}\cup S_{new})$
  \end{description}
\item[{\bf until}] $\pdset{S}_{old}=\pdset{S}_{new}$ (modulo variable renaming)
\item[{\bf output}] $\pdset{S}:=\pdset{S}_{new}$
\end{description}

\end{procedurerm}

The above procedure is parametrised by
 an unfolding rule $\mathit{unfold}$
 and an generalisation
 operation $\mathit{generalise}$.
The latter can be used to ensure termination and
 can be formally defined as follows.

\begin{definition} \label{def:abstract}
An {\em generalisation operation\/} is a function $\mathit{generalise}$ 
 from sets of atoms to sets of atoms
 such that, for any finite set of atoms $S$,
  $\mathit{generalise}(S)$ is a finite set of atoms $S'$ using the
 same predicates as those in $S$, and
 every atom in $S$ is an instance
 of an atom in $S'$.
\end{definition}

\noindent If Procedure~\ref{skeleton:pdalgorithm} terminates then
 the closedness condition is satisfied.
Finally, note that, two sets of atoms $S_1$ and $S_2$ 
 are said to be identical {\em modulo variable renaming\/}
 if for every $s_1\in S_1$
 there exists $s_2\in S_2$ such that $s_1$ and $s_2$ are variants,
 and vice versa.


\subsection{Off-Line Partial Deduction and Binding-Types}
\label{subsection:off-line-pd}

In Procedure~\ref{skeleton:pdalgorithm} one can distinguish
 between two different
 levels of control.
The unfolding rule $U$ controls the construction of the incomplete
 SLDNF-trees.
This is called the {\em local control\/}
 \cite{Gallagher:PEPM93,MartensGallagher:ICLP95}.
The generalisation operation controls the construction of the set
 of atoms for which such SLDNF-trees are built.
We will refer to this aspect as the {\em global control}.

The control problems have been tackled from
 two different angles: the so-called {\em off-line\/} versus
 {\em on-line\/} approaches.
The {\em on-line\/} approach performs
 all the control decisions
 {\em during\/}
 the actual specialisation phase.
The {\em off-line\/} approach  on the other hand
performs an analysis phase {\em prior\/} to
 the actual specialisation phase, based on a
 description of what kinds of specialisations will be required.
This analysis phase provides annotations which then guide the 
 specialisation phase proper,
 often to the point of making it almost trivial.

Partial evaluation of functional programs \cite{ConselDanvy:POPL93,Jones:peval}
has mainly stressed off-line approaches, while supercompilation of functional
\cite{Turchin:toplas86,SorensenGluck:ILPS95}
and partial deduction of logic programs
\cite{Gallagher91:ngc,Sahlin93:ngc,bol:jlp93-pd,Bruynooghe:ngc92,%
MartensDeSchreye:jlp95b,MartensGallagher:ICLP95,%
LeuschelMartensDeSchreye:Toplas,%
CPD:megapaper}
have mainly concentrated on on-line control.

An initial motivation for using the off-line approach was to achieve
 effective self-application \cite{JonesSestoftSondergaard:LSC89}.
But the off-line approach is in
general also much more efficient
 since many decisions
concerning control are made {\em before\/} and not during specialisation. 
This is especially true in a setting where
 the same program is re-specialised several times.
(Note, however, that the global control is 
 usually not done in a fully offline
 fashion: almost all offline 
 partial evaluators maintain during specialisation 
 a list of calls that have been previously specialised or are pending 
 \cite{Jones:peval}.)

Most off-line approaches perform what is called a {\em binding-time analysis
  ({\em BTA})} prior to the specialisation phase.
The purpose of this analysis is to figure out which values will
 be known at specialisation time proper and which values will only
 be known at runtime.
The simplest approach is to classify arguments within the program
 to be specialised as either {\em static} or {\em dynamic}.
The value of a static
 argument will be {\em definitely known\/} (bound) at
 specialisation time whereas a dynamic argument is not necessarily known 
 at specialisation time.
In the context
 of partial deduction of logic programs,
 a static argument can be seen \cite{MogensenBondorf:LOPSTR92}
 as being a term which
 is guaranteed not to be more instantiated at run-time
 (it can never be less instantiated at run-time; otherwise the
  information provided would be incorrect).
For example if we specialise a program for all instances of
 $p(a,X)$ then the first argument to $p$ is static while the second
 one is 
 dynamic%

This approach is successful for functional programs, but often
proves to be too weak for logic programs: in logic programming
partially instantiated data structures appear naturally even at
runtime. A simple classification of arguments into ``fully
known'' or ``totally unknown'' is therefore unsatisfactory and
would prevent specialising a lot of ``natural'' logic programs such
as the vanilla metainterpreter
\cite{HillGallagher:metachapter,MartensDeSchreye:metachapter}
 or most of the benchmarks from the {\sc dppd} library
 \cite{Leuschel96:ecce-dppd}.

The basic idea to improve upon the above shortcoming, is to describe
 parts of arguments which will actually be known at specialisation time
 by a special form of types.%
\footnote{
This is somewhat related to the way instantiations are defined in
 the Mercury language \cite{SomogyiHendersonConway:jlp}.
 But there are major differences, which we discuss later.}
Below, we will develop the first such description, of what we call 
{\em binding-types}, in logic programming.

\subsubsection*{Binding-Types}

In logic programming,
 a type can be defined as a
 set of terms closed under substitution
 \cite{AptMarchiori:ModesTypes}.
We will stick to this view
 and adapt the definitions and concepts
 of \cite{YardeniEtAl:TypeBook} (which mainly follow the Hilog notation
 \cite{ChenKiferWarren:NACLP89}).
 

As is common in polymorphically typed
  languages (e.g.~\cite{SomogyiHendersonConway:jlp}), types are
 are built up from type variables and type constructors
   in much the same way as terms are built-up from ordinary variables 
   and function symbols.
Formally, a {\em type\/} is either a {\em type variable\/} or a
 {\em type constructor\/} of 
  arity $n\geq 0$ applied to $n$ types.
 We presuppose the existence of three 0-ary type constructors:
  {\tt\small static}, {\tt\small dynamic}, and {\tt\small nonvar}.
 These constructors will be given a pre-defined meaning below.
Also, a type which contains no variables is called {\em ground}.

 \begin{definition}
     A {\em type definition\/} for a type constructor $c$ of arity $n$ is of 
     the form
     $$c(V_{1},\ldots,V_{n}) \longrightarrow 
        f_{1}(T_{1}^{1},\ldots,T_{1}^{n_{1}}) ~ ; ~ \ldots ~ ; ~
        f_{k}(T_{k}^{1},\ldots,T_{k}^{n_{k}})$$
 with $k\geq 1, n, n_1,\ldots,n_{k}\geq 0$ and where
 $f_{1},\ldots,f_{k}$ are
 distinct function symbols, $V_{1},\ldots,V_{n}$ are distinct 
 type variables, and $T_{i}^{j}$ are types which only contain type variables in
 $\{V_{1},\ldots,V_{n}\}$.

\noindent
 A {\em type system\/} $\Gamma$
  is a set of type definitions, exactly one for every type 
 constructor $c$ different from {\tt\small static}, {\tt\small dynamic}, and {\tt\small
nonvar}.
 We will refer to the type definition for $c$ in $\Gamma$ by
 $\mathit{Def}_{\Gamma}(c)$.
\end{definition}

From now on we will suppose that the underlying type system
 $\Gamma$ is fixed.
A type system $\Gamma_{1}$, defining a type constructor for parametric lists,
 can be defined as follows:
  $\Gamma_1$ = 
  $\{ \mathit{list}(T) \longrightarrow \mathit{nil} ~;~ \mathit{cons}(T,\mathit{list}(T))\}$.
Using the ASCII notations of Mercury \cite{SomogyiHendersonConway:jlp}
 and using Prolog's list notation, the type system 
 $\Gamma_{1}$ would be written down as follows:

{\small
\begin{verbatim}
:- type list(T) ---> [ ] ; [T | list(T)].
\end{verbatim}}

We define {\em type substitutions\/} to be finite sets of the form
 $\{V_{1}/\tau_{1},\ldots,V_{k}/\tau_{k}\}$, where every $V_{i}$ is 
 a type variable and $\tau_{i}$ a type.
Type substitutions can be applied to types (and type definitions)
 to produce {\em instances\/}
 in exactly the same way as substitutions can be applied to terms.
For example, $\mathit{list}(V)\{V/\mathit{static}\} =
\mathit{list}(\mathit{static})$. A type or type definition
 is called {\em ground\/} if it contains no type variables.

 We now define type judgements  relating terms to types in 
  the underlying type system $\Gamma$.
  
\begin{definition}
{\em Type judgements\/} have the form
  $t : \tau$, where $t$ is a term and $\tau$ is a type,
  and are inductively defined as follows.
\begin{itemize}
\item $t : \mathit{dynamic}$ holds for any term $t$
\item $t : \mathit{static} $ holds for any ground term $t$
\item $t : \mathit{nonvar} $ holds for any non-variable term $t$
\item $f(t_{1},\ldots,t_{n}) : c(\tau'_{1},\ldots,\tau'_{k})$ holds
 if there exists a ground instance
  of the type definition $\mathit{Def}_{\Gamma}(c)$ in the underlying
  type system $\Gamma$ which has the form
  $c(\tau'_{1},\ldots,\tau'_{k}) \longrightarrow \ldots
        f(\tau_{1},\ldots,\tau_{n}) \ldots$ and where
  $t_{i} : \tau_{i}$ for $1\leq i \leq n$.
\end{itemize}
\noindent
We also say that a type $\tau$ is {\em more general\/}
 than another type $\tau'$ iff
 whenever $t : \tau'$ then also $t: \tau$.
\end{definition}

Note that our definitions guarantee that
 types are downwards-closed in the sense that for all terms $t$
  and types $\tau$ we have
 $t:\tau$ $\Rightarrow$ $t\theta:\tau$.

Here are a few examples, using the type system $\Gamma_{1}$ above.
First, we have
 $s(0): \mathit{static}$, $s(0): \mathit{nonvar}$,
and $s(0): \mathit{dynamic}$.
Also,  $s(X): \mathit{nonvar}$,  $s(X): \mathit{dynamic}$
 but not $s(X): \mathit{static}$.
For variables we have
 $X: \mathit{dynamic}$, but neither $X: \mathit{static}$
 nor $X: \mathit{nonvar}$.
A few examples with lists (using Prolog's list notation) are as follows:
 $[\,]: \mathit{list}(\mathit{static})$, ~
$s(0): \mathit{static}$ hence
 $[s(0)]: \mathit{list}(\mathit{static})$, ~
$X: \mathit{dynamic}$ and $Y: \mathit{dynamic}$ hence
 $[X,Y]: \mathit{list}(\mathit{dynamic})$.
Finally, we have, for example, that $\mathit{list}(\mathit{dynamic})$ is
more general than $\mathit{list}(\mathit{static})$.
 
\subsubsection*{Binding-Type Analysis and Classification}

We will now formalise the concept of a 
 {\em binding-type\/} analysis
  (which is an extension of a binding-{\em time\/} analysis,
   as in \cite{JorgensenLeuschel:Cogen}).
For that we first define the concept of a division which 
 assigns types to arguments of predicates.

 \begin{definition} \label{def:division}
A {\em division for a predicate $p$} of arity $n$ is an expression of 
the form $p(\tau_{1},\ldots,\tau_{n})$ where each $\tau_{i}$ is a 
ground type.

\noindent
A {\em division for a program $P$} is a set of
  divisions for predicates in $\mathit{Pred}(P)$, with at most one
 division for any predicate.
When there is no ambiguity about the underlying program $P$ we will 
also often simply refer to a {\em division}.

\noindent
A division is called {\em simple\/} iff it contains only the types
 {\tt\small static} and {\tt\small dynamic}.

\noindent
A division $\Delta$ is called {\em more general\/} than another division 
 $\Delta'$ iff $\forall$ $p(\tau'_{1},\ldots,\tau'_{n})\in\Delta'$
  there exists $p(\tau_{1},\ldots,\tau_{n})\in\Delta$ such that for
   $1\leq i \leq n$
  $\tau_{i}$ is more general than $\tau'_{i}$.
\end{definition}

The fact that divisions only use ground types means that 
 we do not cater for polymorhpic types, although we can
 still use parametric types.
This simplifies
 the remainder of the presentation (mainly
 Definition~\ref{def:gen-delta}) but can probably be lifted.
As can be seen from the above definition, we restrict ourselves
 to monovariant divisions in this paper.
As discussed in \cite{Jones:peval}, a
 way to handle polyvariant
 divisions by a monovariant approach
 is to``invent sufficiently
 many versions of each predicate.''
\condtext{}{
Take for example, a program containing the clauses
\begin{quote}
\noindent
\hspace*{1cm} $p(X,Y) \leftarrow q(X),q(Y)$\\
\hspace*{1cm}   $q(a) \leftarrow$\\
\hspace*{1cm}   $q(f(X)) \leftarrow q(X)$
\end{quote}
and where we have 
$\Delta$ = $\{p(\mathit{static},\mathit{static}),$
 $p(\mathit{static},\mathit{dynamic}),$
 $q(\mathit{static}),q(\mathit{dynamic}) \}$.
We can translate this into
$\Delta'$ = $\{p_1(\mathit{static},\mathit{static}),$
 $p_2(\mathit{static},\mathit{dynamic}),$
 $q_1(\mathit{static}),$ $q_2(\mathit{dynamic}) \}$
 for the transformed program:
\begin{quote}
\noindent
\hspace*{1cm} $p_1(X,Y) \leftarrow q_1(X),q_1(Y)$\\
\hspace*{1cm} $p_2(X,Y) \leftarrow q_1(X),q_2(Y)$\\
\hspace*{1cm} $q_1(a) \leftarrow$\\
\hspace*{1cm} $q_1(f(X)) \leftarrow q_1(X)$\\
\hspace*{1cm} $q_2(a) \leftarrow$\\
\hspace*{1cm} $q_2(f(X)) \leftarrow q_2(X)$
\end{quote}
} 

Now, a 
 {\em binding-type analysis\/} will, given a program $P$ (and some
 description of how $P$ will be specialised), perform a pre-processing
 analysis and return a single {\em division\/} for every predicate
 in $P$ describing the part of the values that will be known
 at specialisation time.
It will also return an {\em annotation\/} which will 
 then guide the local unfolding process of the actual partial deduction.
For the time being, an annotation 
 can simply be seen as a particular unfolding rule ${\cal U}$.
We will return to this in Section~\ref{subsec:particual-off-line-pd}.

We are now in a position to formally define a binding-type analysis
 in the context of (pure) logic programs:

\begin{definition}
A {\em binding-type analysis\/} ({\em BTA}) yields, given a program $P$
 and an arbitrary initial division $\Delta_0$
 for $P$, a couple $({\cal U},\Delta)$ consisting of
 an unfolding rule  ${\cal U}$ and a division $\Delta$ for $P$
 more general than $\Delta_0$. We will call the result of a binding-time
 analysis a {\em binding-type classification\/} ({\em BTC}).
\end{definition}

The purpose of the initial
 division $\Delta_0$ is to give information about how the program
 will be specialised: it specifies what form the initial atom(s)
  (i.e., the ones in ${\cal S}_0$ of Procedure~\ref{skeleton:pdalgorithm})
  can take.
The r\^{o}le of $\Delta$ is to give information
 about the atoms and their binding types
 that can occur at the global level
 (i.e., the ones in ${\cal S}_{new}$ and ${\cal S}_{old}$
  of Procedure~\ref{skeleton:pdalgorithm}).
In that light, not all {\em BTC\/} are
 correct and we have to develop a safety criterion.
Basically a {\em BTC\/} is safe iff every atom that can potentially
appear in one of the sets ${\cal S}_{new}$ 
 of Procedure~\ref{skeleton:pdalgorithm}
(given the restrictions imposed by the annotation of the {\em BTA}) corresponds
to the patterns described by $\Delta$.%
\footnote{Our safety condition
 differs somewhat from the
 classical {\em uniform congruence\/} requirement
 \cite{Launchbury:91:book,Jones:peval}.
We discuss this difference in Section~\ref{section:related-work}.}

We first define a safety notion for atoms and goals.

\begin{definition} 
Let $P$ be a program and let $\Delta$ be a division for $P$ and
 let $p(t_{1},\ldots,t_{n})$ be an atom.
Then $p(t_{1},\ldots,t_{n})$ is {\em safe wrt $\Delta$\/} iff
 $\exists p(\tau_{1},\ldots,\tau_{n})\in\Delta$ such that
 $\forall i\in\{1,\ldots,n\}$ we have
  $t_{i} : \tau_{i}$.
A set of atoms $S$ is {\em safe wrt $\Delta$\/} iff
 every atom in $S$ is safe wrt $\Delta$.
Also a goal $G$ is {\em safe wrt $\Delta$\/} iff all the atoms occurring in $G$
 are safe wrt $\Delta$.
\end{definition}

For example $p(a,X)$ and $\larr p(a,a), p(b,c)$ are
 safe wrt $\Delta = \{p(\mathit{static},\mathit{dynamic})\}$
 while $p(X,a)$ is not.

\begin{definition} 
    \label{def:safety} \label{def:safebta} 
  Let $\beta = ({\cal U},\Delta)$ be a {\em BTC\/} for a program $P$.
  Then $\beta$ is a
  {\em globally safe {\em BTC\/} for $P$\/} iff for every goal $G$ which is safe wrt
  $\Delta$, ${\cal U}(P,G)$ is an SLDNF-tree $\tau$ for
   $P\cup\{G\}$ whose
  leaf goals are safe wrt $\Delta$.
  \noindent
  A {\em BTA\/}
  is {\em globally safe\/} if for any program $P$ it produces a globally
  safe {\em BTC\/} for $P$.
\end{definition}

Sometimes --- in order to simplify both
 the partial deducer and the {\em BTA\/} --- one might want 
to generalise atoms and then lift them to the global level
 (i.e., ${\cal S}_{new}$ in Procedure~\ref{skeleton:pdalgorithm})
 {\em before\/} the full SLDNF-tree
 $\tau$  has been built, namely at the point where a left-to-right
 selection rule would have selected the atom.
This is the motivation behind the following notion of
 a {\em strongly\/} globally safe
 {\em BTC}.

\begin{definition} 
    \label{def:strong-global-safety} 
  Let $\beta = ({\cal U},\Delta)$ be a {\em BTC\/} for a program $P$.
  Then 
 $\beta$ is a {\em strongly globally safe\/}  {\em BTC\/} for $P$
 iff it is globally safe for $P$ and for every goal $G$ which is safe wrt
  $\Delta$, ${\cal U}(P,G)$ is an SLDNF-tree such that
 the literals to the left of selected literals are also safe wrt $\Delta$.
\end{definition}

Notice, that in the above definitions of safety
 no requirement is made about the actual
atoms selected by ${\cal U}$.
Indeed, contrary to functional or imperative programming languages, 
 definite logic programs can handle uninstantiated variables and
 a positive atom can always be selected.
Nonetheless, if we have negative literals or Prolog built-ins, this is no 
longer true. For example, {\tt\small X is Y + 1} can only be selected if 
{\tt\small Y} is ground. Put in other terms, we can only select a call
 ``{\tt\small $s$ is $t$}'' if it is safe wrt $\{is(\mathit{dynamic},\mathit{static})\}$.
Also, we might want to restrict unfolding of user-defined predicates 
 to cases where only one clause matches. For example, we might want to 
unfold a call $\mathit{app}(r,s,t)$ (see Example~\ref{example:append} below)
 only if it is safe wrt
 $\{\mathit{app}(\mathit{static},\mathit{dynamic},\mathit{dynamic})\}$.
This motivates the next definition, which can be used to ensure that 
 only properly instantiated calls to built-ins and atoms are selected.

\begin{definition}
  A {\em BTC\/}
   $\beta = ({\cal U},\Delta)$ is {\em locally safe for $P$\/}
   iff for every goal $G$ which is safe wrt
  $\Delta$, ${\cal U}(P,G)$ is an SLDNF-tree for
   $P\cup\{G\}$ where all
  selected literals are safe wrt $\Delta$.
\end{definition}

\begin{figure}

\setlength{\unitlength}{2300sp}%
\begingroup\makeatletter\ifx\SetFigFont\undefined%
\gdef\SetFigFont#1#2#3#4#5{%
  \reset@font\fontsize{#1}{#2pt}%
  \fontfamily{#3}\fontseries{#4}\fontshape{#5}%
  \selectfont}%
\fi\endgroup%
\begin{picture}(6312,3123)(3601,-5665)
\thicklines
{\put(6151,-4261){\vector( 0,-1){1050}}
}%
{\put(4801,-2836){\vector(-1,-1){1050}}
}%
\thinlines
{\multiput(6751,-5536)(9.00775,0.00000){130}{\makebox(1.6667,11.6667){\SetFigFont{5}{6}{\rmdefault}{\mddefault}{\updefault}.}}
\put(9001,-4410){\vector( 1, 1){0}}
\multiput(7913,-5498)(6.35965,6.35965){171}{\makebox(1.6667,11.6667){\SetFigFont{5}{6}{\rmdefault}{\mddefault}{\updefault}.}}
}%
{\multiput(6226,-3886)(7.46746,4.97830){118}{\makebox(1.6667,11.6667){\SetFigFont{5}{6}{\rmdefault}{\mddefault}{\updefault}.}}
\put(9076,-3286){\vector( 1, 0){0}}
\multiput(7088,-3286)(8.99548,0.00000){221}{\makebox(1.6667,11.6667){\SetFigFont{5}{6}{\rmdefault}{\mddefault}{\updefault}.}}
}%
{\put(9256,-4606){\oval(210,210)[bl]}
\put(9256,-3316){\oval(210,210)[tl]}
\put(9796,-4606){\oval(210,210)[br]}
\put(9796,-3316){\oval(210,210)[tr]}
\multiput(9256,-4711)(9.00000,0.00000){61}{\makebox(1.6667,11.6667){\SetFigFont{5}{6}{\rmdefault}{\mddefault}{\updefault}.}}
\multiput(9256,-3211)(9.00000,0.00000){61}{\makebox(1.6667,11.6667){\SetFigFont{5}{6}{\rmdefault}{\mddefault}{\updefault}.}}
\multiput(9151,-4606)(0.00000,9.02098){144}{\makebox(1.6667,11.6667){\SetFigFont{5}{6}{\rmdefault}{\mddefault}{\updefault}.}}
\multiput(9901,-4606)(0.00000,9.02098){144}{\makebox(1.6667,11.6667){\SetFigFont{5}{6}{\rmdefault}{\mddefault}{\updefault}.}}
}%
{\put(9005,-4071){\vector( 1, 0){0}}
\multiput(7730,-4071)(8.97887,0.00000){142}{\makebox(1.6667,11.6667){\SetFigFont{5}{6}{\rmdefault}{\mddefault}{\updefault}.}}
}%
\thicklines
{\put(4951,-2836){\vector( 1,-1){1050}}
}%
\put(5151,-5611){\makebox(0,0)[lb]{\smash{\SetFigFont{12}{14.4}{\rmdefault}{\mddefault}{\updefault}{\small $\larr p(u1,u2)$}%
}}}
\put(3601,-4111){\makebox(0,0)[lb]{\smash{\SetFigFont{12}{14.4}{\rmdefault}{\mddefault}{\updefault}{\small $\Box$}%
}}}
\put(5531,-4111){\makebox(0,0)[lb]{\smash{\SetFigFont{12}{14.4}{\rmdefault}{\mddefault}{\updefault}{\small $\larr p(t1,t2), \, \underline{q(t1)}$}%
}}}
\put(4031,-2686){\makebox(0,0)[lb]{\smash{\SetFigFont{12}{14.4}{\rmdefault}{\mddefault}{\updefault}{\small $\larr \underline{p(s1,s2)}$}%
}}}
\put(9341,-4061){\makebox(0,0)[lb]{\smash{\SetFigFont{12}{14.4}{\rmdefault}{\mddefault}{\updefault}{$\Delta$}%
}}}
\put(7200,-3161){\makebox(0,0)[lb]{\smash{\SetFigFont{12}{14.4}{\rmdefault}{\mddefault}{\updefault}{\footnotesize strong global}%
}}}
\put(8021,-3931){\makebox(0,0)[lb]{\smash{\SetFigFont{12}{14.4}{\rmdefault}{\mddefault}{\updefault}{\footnotesize local}%
}}}
\put(8276,-5411){\makebox(0,0)[lb]{\smash{\SetFigFont{12}{14.4}{\rmdefault}{\mddefault}{\updefault}{\footnotesize global}%
}}}
\end{picture}
\caption{Different types of safety for a sample, incomplete SLD-tree\label{fig:safety}}
\end{figure}
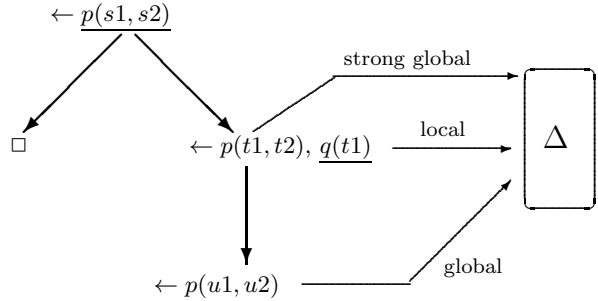

The difference between local and global safety is illustrated in
 Figure~\ref{fig:safety}.
Note that it might make sense to use
 different divisions for local and global safety.
This can be easily allowed, but we will not do so in the presentation
 of this article.

Let us now return to the global control.
Definition~\ref{def:safebta} requires atoms to be safe in
 the leaves of incomplete SLDNF-trees, i.e.\ at the point where
 the atoms get abstracted and then
 lifted to the {\em global\/} level.
So, in order for Definition~\ref{def:safebta} to ensure safety at
 all stages of Procedure~\ref{skeleton:pdalgorithm},
 the particular generalisation operation employed should not abstract atoms
 which are safe wrt $\Delta$ into atoms which are no longer
 safe wrt $\Delta$.
 
This motivates the following definition:

\begin{definition} \label{def:abstract-safe}
An generalisation operation $\mathit{generalise}$ is {\em safe wrt a
 division $\Delta$} iff
 for every finite set of atoms $S$ which is safe wrt $\Delta$,
 $\mathit{generalise}(S)$ is also safe wrt $\Delta$ .
\end{definition}

In particular this means that
 $\mathit{generalise}$ can only generalise positions marked as {\tt\small dynamic} or the
 arguments of positions marked as {\tt\small nonvar} within the 
 respective binding-type.
For example,
 $\mathit{generalise}(\{p([\,])\}) = \{p(X)\}$ is neither safe wrt
  $\Delta = \{p(\mathit{static})\}$ nor wrt
  $\Delta' = \{p(\mathit{nonvar})\}$ nor wrt
  $\Delta'' = \{p(\mathit{list(dynamic)})\}$, but it is safe wrt
 $\Delta''' = \{p(\mathit{dynamic})\}$.
Also, $\mathit{generalise}(\{p(f([\,]))\}) = \{p(f(X))\}$ is not safe wrt
 $\Delta = \{p(\mathit{static})\}$ but is safe wrt both
 $\Delta' = \{p(\mathit{nonvar})\}$ and
$\Delta''' = \{p(\mathit{dynamic})\}$.


\begin{example}\label{example:append}
Let  $P$ be the well known append program
\begin{quote}
\noindent
\hspace*{1cm} $\mathit{app}([\,],L,L) \leftarrow$\\
\hspace*{1cm} $\mathit{app}([H|X],Y,[H|Z]) \leftarrow \mathit{app}(X,Y,Z)$
\end{quote}
Let $\Delta = \{\mathit{app}(\mathit{static},\mathit{dynamic},\mathit{dynamic}) \}$ and let
 ${\cal U}$ be any unfolding rule.
Then $({\cal U},\Delta)$ is a globally and locally safe {\em BTC\/} for $P$.
E.g., the goal $\leftarrow \mathit{app}([a,b],Y,Z)$ is safe
 wrt $\Delta$ and $\cal U$ can either stop at
 $\leftarrow \mathit{app}([b],Y,Z)$,
 $\leftarrow \mathit{app}([\,],Y',Z')$ or at the empty goal $\Box$.
All of these goals are safe wrt $\Delta$.
More generally, unfolding a goal $\leftarrow  \mathit{app}(t_{1},t_{2},t_{3})$
 where $t_{1}$ is ground (and thus static),
 leads only to goals whose first arguments are ground (static).
\end{example}


\subsection{{\sc lix}, a Particular Off-Line Partial Deduction Method}
\label{subsec:particual-off-line-pd}

In this subsection we define a specific off-line partial deduction method
which will serve as the basis for the $\mathit{cogen}$
developed in the remainder of
this paper.  For simplicity, we will, until further notice,
 restrict ourselves to
definite programs. Negation will in practice be treated
in the $\mathit{cogen}$ either
as a built-in or via the {\em if-then-else} construct 
 (both of which we will discuss later).
 
We first define a particular class of simple-minded but 
effective unfolding rules.

\begin{definition} 
    \label{def:annotation} 
An {\em annotation\/} ${\cal A}$ for a program $P$
 marks every literal in the body of each clause of $P$ 
 as either {\em reducible\/} or {\em non-reducible}.
A program $P$ together with an annotation ${\cal A}$ for $P$ is called
 an {\em annotated program}, and is denoted by $P_{{\cal A}}$.
 
 \noindent
Given an annotation ${\cal A}$ for $P$,
 $U_{{\cal A}}$ denotes the unfolding rule which given a goal $G$
 computes
 $U_{{\cal A}}(P,G)$ by
 unfolding the leftmost atom in $G$ and then
 continously unfolds leftmost
 reducible atoms until an SLD-tree is
 obtained with only non-reducible atoms in the leaves.
\end{definition}

Syntactically we represent an annotation for $P$ by underlining
 the predicate symbol of reducible literals.%
\footnote{In functional programming one usually underlines the
 non-reducible calls. But in logic programming underlining
 a literal is usually used to denote selected literals
 and therefore underlining
 the reducible calls is more intuitive.}

\begin{example}
Let  $P_{{\cal A}}$ be the following annotated program
\begin{quote}
\noindent
\hspace*{1cm}
   $p(X) \leftarrow \underline{q}(X,Y),\underline{q}(Y,Z)$\\
\hspace*{1cm}   $q(a,b) \leftarrow$\\
\hspace*{1cm}   $q(b,a) \leftarrow$
\end{quote}
Let $\Delta$ = $\{p(\mathit{static}), q(\mathit{static},\mathit{dynamic}) \}$.
Then $\beta$ = $(U_{{\cal A}},\Delta)$ is a globally
 safe {\em BTC\/} for $P$.
For example the goal $\leftarrow p(a)$ is safe
 wrt $\Delta$ and unfolding it according to $U_{{\cal A}}$
 will lead
 (via the intermediate goals $\leftarrow q(a,Y),q(Y,Z)$ and 
        $\leftarrow q(b,Z)$)
  to the empty goal $\Box$ which is safe wrt $\Delta$.
Note that every selected atom is safe wrt $\Delta$,
 hence $\beta$ is actually also locally safe for $P$.
Also note that $\beta' = (U_{{\cal A}'},\Delta)$,
 where ${\cal A}'$ marks every literal as non-reducible, is 
{\em not\/}
 a safe {\em BTC\/} for $P$.
For instance, given the goal $\leftarrow p(a)$
 the unfolding rule $U_{{\cal A}'}$ just performs
 one unfolding step and thus
 stops at the goal $\leftarrow q(a,Y),q(Y,Z)$ which
 contains the unsafe atom $q(Y,Z)$.
\end{example}

From now on we will only use 
 unfolding rules of the form $U_{{\cal A}}$ obtained from an annotation
 ${\cal A}$ and
 our {\em BTAs} will thus return results of the form
        $\beta = (U_{{\cal A}},\Delta)$.

Given we have a BTC for a program $P$,
 in order to arrive at a concrete
 instance of Procedure~\ref{skeleton:pdalgorithm} we now only need
 a (safe) generalisation operation, which we define in the following.

\begin{definition} 
    \label{def:gen-delta}
We first define a family of
 mappings $\mathit{gen}_{\tau}$ from terms to terms, parameterised by 
 types,
  inductiely as follows:
 \begin{itemize}
     \item $\mathit{gen}_{\mathit{static}}(t) = t$, for any term $t$
     \item $\mathit{gen}_{\mathit{dynamic}}(t) = V$, for any term $t$
        and where $V$ is a fresh variable
     \item $\mathit{gen}_{\mathit{nonvar}}(f(t_{1},\ldots,t_{n}))$ = 
                $f(V_{1},\ldots,V_{n})$, ~
        where $V_{1},\ldots,V_{n}$ are $n$ distinct fresh variables
     \item $\mathit{gen}_{c(\tau'_{1},\ldots,\tau'_{k})}(f(t_{1},\ldots,t_{n}))$ = 
       $f(\mathit{gen}_{\tau_{1}}(t_{1}),\ldots,\mathit{gen}_{\tau_{n}}(t_{n}))$,  
       if there exists
        a ground instance in
        $\mathit{Def}_{\Gamma}(c)$ of the form
  $c(\tau'_{1},\ldots,\tau'_{k}) \longrightarrow \ldots ;
        f(\tau_{1},\ldots,\tau_{n})  ; \ldots$.
\end{itemize}
       
\noindent
Let $\Delta$ be a division for some program $P$.
We then define the partial mapping $\mathit{gen}_{\Delta}$ from atoms
 to atoms by:
\begin{itemize}
\item
 $\mathit{gen}_{\Delta}(p(t_{1},\ldots,t_{n})) = 
  p(\mathit{gen}_{\tau_{1}}(t_{1}),\ldots,\mathit{gen}_{\tau_{n}}(t_{n}))$
  if $\exists$ $p(\tau_{1},\ldots,\tau_{n})$ $\in\Delta$
  such that $p(t_1,\ldots,t_n) : p(\tau_{1},\ldots,\tau_{n})$.
\end{itemize}

\noindent
We also define the
 generalisation operation $\mathit{generalise}_{\Delta}$ as follows: 
For a set $S$ of atoms which is safe w.r.t.\ $\Delta$,
  $\mathit{generalise}_{\Delta}(S)$ is 
  a minimal subset $S_1$ of
    $S_2 = \{\mathit{gen}_{\Delta}(s) \mid s\in S\}$ 
    such that for every element $s$ of $S_2$ there exists
    a variant of $s$ in $S_1$.
\end{definition}

For example, if 
 $\Delta$ = $\{ p(\mathit{static},\mathit{dynamic}),$
  $q(\mathit{dynamic},\mathit{static},\mathit{nonvar})\}$ we have
$\mathit{gen}_{\Delta}(p(a,b))$ = $p(a,X)$ and $\mathit{gen}_{\Delta}(q(a,b,f(c)))$ =
$q(Y,b,f(Z))$. We also have that
 $\mathit{generalise}_\Delta(\{p(a,b),$ $q(a,b,f(c))\})$ = $\{p(a,X), q(Y,b,f(Z))\}$.
\\
For $\Delta'$ = $\{ r(\mathit{list(dynamic)}) \}$
 (where $\mathit{list(dynamic)}$ is 
defined in Section~\ref{subsection:off-line-pd}) we have that
 $\mathit{gen}_{\Delta}(r([a,b,c]))$ = $r([X,Y,Z])$ and
 $\mathit{gen}_{\Delta}(r([H|T]))$ is undefined
 because it is not safe w.rt.\ $\Delta$.


As can be seen $\mathit{gen}_{\Delta}$ is in general not total,
 but is total for atoms safe wrt $\Delta$.
Hence, in the context of a globally safe {\em BTA},
 $\mathit{gen}_{\Delta}$ and $\mathit{generalise}_{\Delta}$ will always be defined.

\begin{proposition}
For every division $\Delta$, $\mathit{generalise}_{\Delta}$ is safe wrt $\Delta$.
\end{proposition}

\ignore{ ***** Re-use somewhere else ??
Note that in Procedure~\ref{skeleton:pdalgorithm}
 the atoms in $\mathit{leaves}(unfold(P,s_{o}))$ are all added and abstracted
 simultaneously, i.e.\ the procedure progresses in a breadth-first manner.
In general this will yield a different result from a depth-first
 progression (i.e.\ adding one atom at a time).
However, due to its simplicity,
 $\mathit{generalise}_{\Delta}$ is a homomorphism%
\footnote{I.e.\ $\mathit{generalise}(\emptyset) = \emptyset$ and
        $\mathit{generalise}(S\cup S') = \mathit{generalise}(S) \cup
\mathit{generalise}(S')$.}
        (up to variable renaming)
 and thus
 both progressions will yield exactly the same set of atoms and thus the
 same specialisation.
This is something which we will actually exploit in the
 practical implementation to make use of a depth-first 
 progression.
***** }

Based upon this generalisation operation, we can also define a 
corresponding renaming and filtering operation:

\begin{definition}\label{def:filter-delta}
 Let $\|.\|$ be a fixed mapping from atoms to natural numbers such that
  $\|A\| = \|B\|$ iff $A$ and $B$ are variants.
We then define $\mathit{filter}_{\Delta}$ as follows:
 $\mathit{filter}_{\Delta}(A)$ = 
   $p_{\|\mathit{gen}_{\Delta}(A)\|}(V_{1}\theta,\ldots,V_{k}\theta)$,
   where
    $A = \mathit{gen}_{\Delta}(A)\theta$,
    $p$ is the predicate symbol of $A$, and
    $V_{1},\ldots,V_{k}$ are the variables appearing in
  $\mathit{gen}_{\Delta}(A)$.
\end{definition}

The purpose of the mapping $\|.\|$ is to assign
to every specialised atom
 (i.e., atoms of the form $\mathit{gen}_{\Delta}(A)$)
 a unique identifier and predicate name, thus ensuring
  the independence condition \cite{Lloyd:jlp91}.
The $\mathit{filter}_{\Delta}$ operation will properly rename instances of these atoms
 and also filter out static parts, thus improving the 
efficiency of the residual code
 \cite{Gallagher:META90,Benkerimi93:jlc}.
For example, given the division
 $\Delta$ = $\{ p(\mathit{static},\mathit{dynamic}),$
$q(\mathit{dynamic},\mathit{static},\mathit{nonvar})\}$,
 $\|p(a,X)\| = 1$,
 and $\|q(X,b,f(Y)\| = 2$
we have that
$\mathit{filter}_{\Delta}(p(a,b))$ = $p_{1}(b)$
as well as $\mathit{filter}_{\Delta}(q(a,b,f(c)))$ = $q_{2}(a,c)$.

In the remainder of this paper we will use 
 the following off-line partial deduction method: 
\begin{procedure}{\vspace*{-0.5cm}~~~~~~~~~~~~~~~~~~(off-line partial
deduction)}\label{algo:bta-pd}
\begin{enumerate}
\item Perform a globally safe {\em BTA\/} (possibly by hand) returning results of the form 
 $(U_{{\cal A}},\Delta)$.
\item Perform Procedure~\ref{skeleton:pdalgorithm} with $U_{{\cal A}}$ as
 unfolding rule and $\mathit{generalise}_{\Delta}$ as generalisation operation.
 The initial set of atoms ${\cal S}_{0}$ should only contain atoms which
 are safe wrt $\Delta$.
\item Construct the 
 specialised program $P'$ using $\mathit{filter}_{\Delta}$ and the
  output ${\cal S}$ of Procedure~\ref{skeleton:pdalgorithm} 
  as follows:
 $P'$ =
  $\{ \mathit{filter}_{\Delta}(A)\theta \leftarrow$ 
     $\mathit{filter}_{\Delta}(B_1),\ldots, \mathit{filter}_{\Delta}(B_n)$
     $\mid$
     $A\theta \leftarrow B_1,\ldots,B_n \in \mathit{resultants}(U_{{\cal A}}(P,A))$
     $\wedge$
     $A\in{\cal S} \}$.
\end{enumerate}
\end{procedure}

\begin{proposition}
Let $(U_{{\cal A}},\Delta)$ be a globally safe {\em BTC\/} for a program $P$.
Let ${\cal S}$ be a set of atoms safe wrt $\Delta$.
Then all sets ${\cal S}_{new}$ and ${\cal S}_{old}$
 arising during the execution of
 Procedure~\ref{algo:bta-pd} are safe wrt $\Delta$.
\end{proposition}

Notably, if Procedure~\ref{algo:bta-pd} terminates then the final set 
${\cal S}$ will be safe wrt $\Delta$.
However, none of our notions of safety
 actually ensure (local or global) termination
 of Procedure~\ref{algo:bta-pd}.
Termination is thus another issue (orthogonal to safety)
 which a {\em BTA\/} has to worry about.
Basically, the annotation ${\cal A}$ has to be such
 that for all atoms $A$ which are safe wrt
  $\Delta$, $U_{{\cal A}}$ returns a
 finite SLDNF-tree $\tau$ for
   $P\cup\{\leftarrow A\}$.
Furthermore,
 $\Delta$ has to be such that
 $\mathit{generalise}_{\Delta}$ ensures that only 
 finitely many atoms can appear at the global level.
We will return to this issue
 in Section~\ref{section:bta}.

We now illustrate Procedure~\ref{algo:bta-pd}
 on a relatively simple example.

\begin{example}
\label{example:parser}
We use a small generic parser for a set of languages
 which are defined by grammars of the form $N ::= aN | X$
 (where $a$ is a terminal symbol and
 $X$ is a placeholder for a terminal symbol).
The example is adapted from \cite{komorowski:meta92} and the 
 (annotated) parser $P$
 is depicted in Figure~\ref{figure:parser}.
The first argument to $\mathit{nont}$ is the value for $X$
 while the other two arguments represent the string to be parsed
 as a difference list.

\begin{enumerate}
\item
Given the initial division
         $\Delta_{0} =$ $\{\mathit{nont}(\mathit{static},\mathit{dynamic},\mathit{dynamic})\}$,
 a {\em BTA\/} might return 
 $\beta = (U_{{\cal A}},\Delta)$
         with $\Delta = $
$\{\mathit{nont}(\mathit{static},\mathit{dynamic},\mathit{dynamic}),$ 
       $t(\mathit{static},$ $\mathit{dynamic},\mathit{dynamic})\}$
 and where ${\cal A}$ is represented in Figure~\ref{figure:parser}.
It can be seen that $\beta$ is a globally and locally safe
 {\em BTC\/} for $P$.

\item
Let us now perform the proper partial deduction for
 ${\cal S}_{0} = \{\mathit{nont}(c,T,R)\}$.
Note that the atom $\mathit{nont}(c,T,R)$ is safe wrt $\Delta_{0}$ 
 (and hence also wrt $\Delta$).
Unfolding the atom in
 ${\cal S}_{0}$ yields the SLD-tree in Fig.~\ref{figure:parser-unfold}.
We see that the 
only atom
in the leaves 
is $\{\mathit{nont}(c,V,R)\}$ and we obtain
 ${\cal S}_{old}={\cal S}_{new}$ (modulo variable renaming).
\item
The specialised program before and after filtering is depicted in 
Figure~\ref{figure:parser-spec}.
Note that, if one wishes to call the filtered version in exactly the same way
 as the unfiltered one has to add the clause
$\mathit{nont}(c,T,R) \leftarrow \mathit{nont}_1(T,R)$.
\end{enumerate}
\end{example}

\begin{figure}[thb]
\begin{center}
\renewcommand{\arraystretch}{\mikearraystretch}
\begin{tabular}{@{}l} 
~$\mathit{nont}(X,T,R) \leftarrow$~
 $\underline{t}(a,T,V),\mathit{nont}(X,V,R)$~\\
~$\mathit{nont}(X,T,R) \leftarrow$~
  $\underline{t}(X,T,R)$~\\
~$t(X,[X|R],R) \leftarrow$~\\
\end{tabular}
\caption{A very simple parser \label{figure:parser}}
\end{center}
\end{figure}

\begin{figure}[thb]
\begin{center}
\setlength{\unitlength}{0.009in}%
\begin{picture}(182,132)(133,685)
\thicklines
\put(250,800){\vector( 1,-1){ 30}}
\put(230,800){\vector(-1,-1){ 30}}
\put(190,750){\vector( 0,-1){ 40}}
\put(290,750){\vector( 0,-1){ 40}}
\put(290,785){\makebox(0,0)[rb]{\raisebox{0pt}[0pt][0pt]{\scriptsize ~}}}
\put(180,758){\makebox(0,0)[b]{\raisebox{0pt}[0pt][0pt]{\footnotesize
$\leftarrow \underline{t(a,T,V)},\mathit{nont}(c,V,R)$}}}
\put(295,758){\makebox(0,0)[b]{\raisebox{0pt}[0pt][0pt]{\footnotesize
$\leftarrow \underline{t(c,T,R)}$}}}
\put(190,698){\makebox(0,0)[b]{\raisebox{0pt}[0pt][0pt]{\footnotesize
$\leftarrow \mathit{nont}(c,V,R)$}}}
\put(290,698){\makebox(0,0)[b]{\raisebox{0pt}[0pt][0pt]{\footnotesize
$\Box$}}}
\put(345,730){\makebox(0,0)[rb]{\raisebox{0pt}[0pt][0pt]{\scriptsize 
{\tt T = [c|R]}}}}
\put(185,730){\makebox(0,0)[rb]{\raisebox{0pt}[0pt][0pt]{\scriptsize 
{\tt T = [a|V]}}}}
\put(205,785){\makebox(0,0)[rb]{\raisebox{0pt}[0pt][0pt]{\scriptsize ~}}}
\put(240,810){\makebox(0,0)[b]{\raisebox{0pt}[0pt][0pt]{\footnotesize
 $\leftarrow \underline{\mathit{nont}(c,T,R)}$}}}
\end{picture}
\caption{Unfolding the parser of Figure~\protect\ref{figure:parser}
 \label{figure:parser-unfold}}
\end{center}
\end{figure}
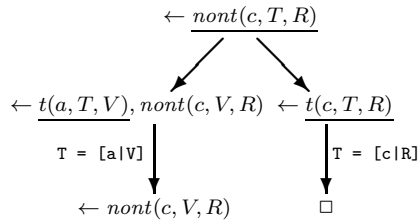

\begin{figure}[thb]
\begin{center}
\renewcommand{\arraystretch}{\mikearraystretch}
\begin{tabular}{@{}l}  
~ $\mathit{nont}(c,[a|V],R) \leftarrow \mathit{nont}(c,V,R)$\\
~ $\mathit{nont}(c,[c|R],R) \leftarrow$\\
\end{tabular}

~\\[1ex]
\begin{tabular}{@{}l} 
~ $\mathit{nont}_1([a|V],R) \leftarrow \mathit{nont}_1(V,R)$\\
~ $\mathit{nont}_1([c|R],R) \leftarrow$\\
\end{tabular}
\caption{Unfiltered and filtered specialisation of
 Figure~\protect\ref{figure:parser} \label{figure:parser-spec}}
\end{center}
\end{figure}


\subsubsection*{The {\sc lix} system}
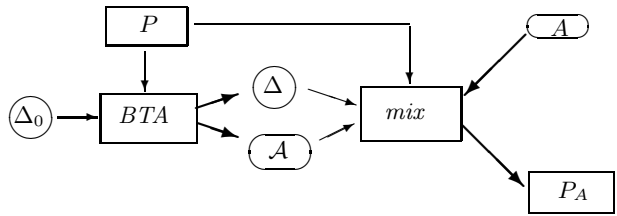
\begin{figure}[tb]
\begin{center}\begin{small}
\setlength{\unitlength}{0.72pt}%
\begin{picture}(351,134)
\thicklines
              \put(110,63){\vector(3,-1){21}}
              \put(109,71){\vector(3,1){22}}
              \put(249,62){\vector(1,-1){32}}
              \put(282,110){\vector(-1,-1){32}}
\thinlines    \put(299,24){$P_{A}$}
              \put(284,16){\framebox(44,21){}}
              \put(295,109){$A$}
              \put(299,114){\oval(32,12)}
              \put(221,114){\vector(0,-1){30}}
              \put(108,115){\line(1,0){113}}
              \put(82,101){\vector(0,-1){21}}
              \put(79,111){$P$}
              \put(62,104){\framebox(42,20){}}
              \put(13,63){$\Delta_{0}$}
              \put(147,45){${\cal A}$}
              \put(144,79){$\Delta$}
              \put(208,65){$\mathit{mix}$}
              \put(68,63){{\em BTA\/}}
              \put(174,54){\vector(2,1){18}}
              \put(168,81){\vector(3,-1){25}}
              \put(196,54){\framebox(52,28){}}
              \put(153,48){\oval(32,18)}
              \put(36,66){\vector(1,0){21}}
              \put(150,82){\circle{24}}
              \put(59,53){\framebox(50,25){}}
              \put(22,66){\circle{24}}
\end{picture}
\end{small}
\caption{Overview of the $\mathit{mix}$ approach
 \label{figure:overview-mix}}
\end{center}
\end{figure}

Based upon Procedure~\ref{algo:bta-pd}
 we have implemented a concrete offline partial deduction
 system called {\sc lix} using the
 traditional $\mathit{mix}$ approach
 \cite{Jones:peval} depicted in Figure~\ref{figure:overview-mix}.
We will examine the power of this system
 in more detail in Section~\ref{section:results}.
As we will see, provided that a good $\mathit{BTA}$ is used,
 the quality of the specialised code provided by
 {\sc lix} can be surprisingly good.
As is to be expected, due to its offline nature,
  {\sc lix} itself is very fast.
In the next section, we show how the specialisation
 speed can be further improved by using the $\mathit{cogen}$
 approach.

Now, a crucial aspect for the performance of {\sc lix} is of course
 the quality of the $\mathit{BTC}$.
Also, the runtime of an automatic $\mathit{BTA}$ can 
 usually not be neglected, and it could be considerably higher than
 that of {\sc lix}.
However, in cases where the same code is specialised over and over
 again, the cost of the $\mathit{BTA}$ is much less significant,
 as it only has to be run once.
We will return to these issues in
 Sections~\ref{section:results} and \ref{section:bta}.

\section{The $\mathit{cogen}$ approach for logic programming}
\label{section:cogen-pure}


Based upon the generic offline partial deduction framework presented 
in the previous section, we will now describe the $\mathit{cogen}$
 approach to logic program specialisation.
 
 \subsection{General Overview}
 
In the context of our framework, a
 {\em generating extension\/} for a program $P$ wrt to a given
 safe {\em BTC\/} $(U_{{\cal A}},\Delta)$ for $P$, is a program that
 receives as its only input an atom $A$ which is
 safe wrt $\Delta$,
 which it then specialises
 (using parts 2 and 3 of Procedure~\ref{algo:bta-pd}
  with ${\cal S}_0 = \{A\}$),
thereby producing a specialised program 
 $P_{A}$. 
In the particular context of
 Example~\ref{example:parser} a generating extension is a program that, when
 given the safe atom $\mathit{nont}(c,T,R)$, produces the residual program shown 
 in Figure~\ref{figure:parser-spec}.

In this section, we develop the
 {\em compiler generator\/} {\sc logen}; it is a program that given a
program $P$ and a globally safe {\em BTC\/} $\beta = (U_{{\cal A}},\Delta)$
 for $P$, produces a
 generating extension for $P$ wrt $\beta$.

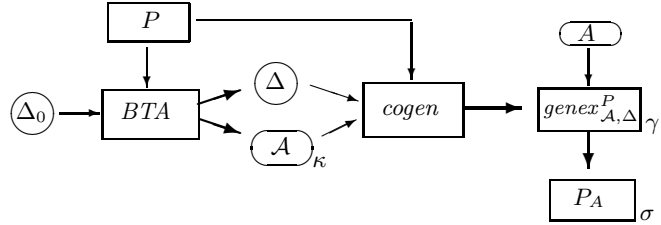
\begin{figure}[tb]
\begin{center}\begin{small}
\setlength{\unitlength}{0.72pt}%
\begin{picture}(351,134)
\thicklines
              \put(110,63){\vector(3,-1){21}}
              \put(109,71){\vector(3,1){22}}
              \put(249,69){\vector(1,0){30}}
              \put(314,57){\vector(0,-1){20}}
\thinlines    \put(306,18){$P_{A}$}
              \put(341,10){$\sigma$}
              \put(308,104){$A$}
              \put(294,10){\framebox(44,21){}}
              \put(314,100){\vector(0,-1){19}}
              \put(315,108){\oval(32,12)}
              \put(221,114){\vector(0,-1){30}}
              \put(108,115){\line(1,0){113}}
              \put(82,101){\vector(0,-1){21}}
              \put(79,111){$P$}
              \put(62,104){\framebox(42,20){}}
              \put(13,63){$\Delta_{0}$}
              \put(147,45){${\cal A}$}
              \put(169,38){$\kappa$}
              \put(144,79){$\Delta$}
              \put(290,66){{\footnotesize $\mathit{genex}^{P}_{{\cal A},\Delta}$}}
              \put(344,58){$\gamma$}
              \put(205,65){$\mathit{cogen}$}
              \put(68,63){{\em BTA\/}}
              \put(174,54){\vector(2,1){18}}
              \put(168,81){\vector(3,-1){25}}
              \put(288,57){\framebox(53,22){}}
              \put(196,54){\framebox(52,28){}}
              \put(153,48){\oval(32,18)}
              \put(36,66){\vector(1,0){21}}
              \put(150,82){\circle{24}}
              \put(59,53){\framebox(50,25){}}
              \put(22,66){\circle{24}}
\end{picture}\end{small}
\caption{Overview of the $\mathit{cogen}$ approach
 \label{figure:overview}}
\end{center}
\end{figure}

An overview of the whole process is depicted in
 Figure~\ref{figure:overview} (the $\kappa$, $\gamma$, and $\sigma$ 
 subscripts will be explained in the next section), and
 also shows the differences with the more traditional
 $\mathit{mix}$ approach presented in Figure~\ref{figure:overview-mix}.
As can be seen, $P$, $\Delta$, and ${\cal A}$ have been compiled into 
the generating extension $\mathit{genex}^{P}_{{\cal A},\Delta}$
 (contributing to its efficiency and also making it standalone).
A generating extension is thus not a generic partial 
 evaluator, but a highly specialised one: it can 
 specialise the program $P$  only 
 for calls $A$ which are safe wrt $\Delta$ and it can 
 only follow the annotation ${\cal A}$.

To explain and formalise the $\mathit{cogen}$ approach, we
 will first examine the r\^{o}le and structure of
 generating extensions
 $\mathit{genex}^{P}_{{\cal A},\Delta}$.
Once this is clear we will consider
 how the {\em cogen\/} can generate them.


 \subsection{The local control} \label{section:cogen:local}

The crucial idea for simplicity (and efficiency) of the generating
 extensions is to produce a specific ``unfolding'' predicate $p_u$
 for each predicate $p/n$.
Also, for every predicate which is susceptible to appear at the
 global level, we will produce a specific ``memoisation'' predicate $p_m$.

\begin{figure}[htb]
\begin{center}\begin{small}
  \setlength{\unitlength}{0.4pt}
\begin{picture}(344,120)
\thinlines    \put(-35,63){$U_{{\cal A}}$:}
              \put(246,66){$\theta$ $\cup$ {\footnotesize $\{B/[L_{1},...,L_{m}]\}$}}
              \put(243,44){{\footnotesize ~}}
              \put(197,101){$p_{u}(t_{1},...,t_{n},B)$}
              \put(225,9){$\Box$}
              \put(233,87){\vector(0,-1){60}}
              \put(25,62){$\theta$}
              \put(135,62){$\specialise_{u}$}
              \put( 9,14){$L_{1},...,L_{m}$}
              \put(47,91){\vector(0,-1){60}}
              \put( 2,101){$p(t_{1},...,t_{n})$}
\end{picture}\end{small}
\caption{Going from $p$ to $p_{u}$
 \label{figure:unfolder-predicate}}
\end{center}
\end{figure}
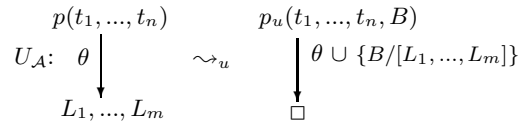

Let us first consider the local control aspect.
This predicate $p_u$ has $n+1$ arguments and is
 tailored towards unfolding calls to $p/n$.
The first $n$ arguments correspond to the arguments of the call
 to $p/n$ which has to be unfolded.
The last argument will collect the
 result of the unfolding process.
More precisely, $p_u(t_1,...,t_n,B)$ will
 succeed for each branch of the incomplete SLDNF-tree obtained by applying
 the unfolding rule
 $U_{{\cal A}}$ to $p(t_1,...,t_n)$, whereby it will return in
 $B$ the atoms in the leaf of the branch
 and also instantiate $t_1,...,t_n$ via the composition of
 $\mathit{mgu}$s of the branch (see Figure~\ref{figure:unfolder-predicate}).
For atoms which get fully unfolded, the
 above can be obtained very {\em efficiently\/} by 
 simply executing the original predicate definition of $p$
 for the goal $\leftarrow p(t_1,...,t_n)$
 (no atoms in the leaves have to be returned because there are none).
To handle the case of incomplete SLDNF-trees we just have to adapt
 the definition of $p$ so that unfolding of non-reducible atoms
  can be prevented
  and the corresponding leaf atoms can be collected in the last argument 
  $B$.

All this can be obtained by transforming every clause for
  $p/n$ into a clause for $p_u/(n+1)$ in the following manner.
To simplify the presentation, we from now on use the
 notation $p(\seq{t})$ to represent an
 atom of the form $p(t_{1},\ldots,t_{n})$
 and also $p(\seq{s},t)$ to represent an
 atom of the form $p(s_{1},\ldots,s_{n},t)$.

We first define the ternary relation
$\kappa \specialise \gamma$ : $\sigma$.
Intuitively (see Figure~\ref{figure:overview}), 
$\kappa$ $\specialise$ $\gamma$ : $\sigma$
    denotes that
   the $\mathit{cogen}$ will produce from the annotated literal or conjunction
   $\kappa$ in the original program $P$
   the calls $\gamma$ in the generating extension $\mathit{genex}^{P}_{{\cal A},\Delta}$.
In turn, the computed answers of  $\gamma$ will instantiate
  $\sigma$ to the bodies of the
 residual clauses that are part of the specialised program
 $P_S$.
If $\gamma$ fails then no residual clause will be
produced. On the other hand, if $\gamma$ has several computed answers then 
 several residual clauses will be produced.
 
\begin{definition} \label{def:pul-transf}
The ternary relation $\kappa \specialise \gamma$ : $\sigma$,
 with $\kappa$ denoting annotated conjunctions,
  $\gamma$ denoting conjunctions
 and $\sigma$ denoting terms,
 is defined by the following three rules. 
Remember that an underlined literal is selected for unfolding.
$$
\begin{array}{ll}
    \underline{p}(\seq{t}) \specialise p_u(\seq{t},C) : C & 
    \mbox{(C fresh variable)}\\[3ex]
    p(\seq{t}) \specialise p_m(\seq{t},C) : C & 
    \mbox{(C fresh variable)}
    \\[3ex]
    \begin{array}{c}
      \forall i: \kappa_{i} \specialise \gamma_{i} : \sigma_{i} \\
      \hline
      (\kappa_{1}, \ldots , \kappa_{n}) \specialise 
      (\gamma_{1}, \ldots , \gamma_{n}) : (\sigma_{1}, \ldots , 
      \sigma_{n})
    \end{array} & \mbox{(conjunctions)}
    \\[2ex]
\end{array}
$$

The above relation can now be used to define the relation $\leadsto_u$
which transforms a clause of $p$ into a clause  for the efficient {\em
  unfolder} $p_u$.

$$
\begin{array}{ll}
    \begin{array}{c}
      p(\seq{t})\leftarrow ~~ \specialise_{u} ~~
       p_{u}(\seq{t},\mathit{true})\leftarrow
    \end{array} & \mbox{(facts)}\\[3ex]
    \begin{array}{c}
      \kappa \specialise \gamma : \sigma \\
      \hline
      p(\seq{t})\leftarrow \kappa ~~ \specialise_{u} ~~
       p_{u}(\seq{t},\sigma)\leftarrow \gamma
    \end{array} & \mbox{(rules)}
    \\[2ex]
\end{array}
$$

Given an annotation ${\cal A}$ and a program $P$
 we define $P_{u}^{\cal A}$
  = $\{c' \mid c\in P \wedge c \specialise_{u} c'\}$.
\end{definition}

Note that
 the transformation $\specialise_{u}$, by means of
 the $\specialise$ transformation, also generates
 calls to $p_{m}$ predicates which we define later.
These predicates
 take care of the global control and also return a filtered
 and renamed version of the call to be specialised as their last argument.

In the above definition
 inserting a literal of the form
 $p_u(\seq{t},C)$ corresponds to further {\em unfolding\/}
 whereas inserting $p_m(\seq{t},C)$ corresponds to stopping local
unfolding
 and leaving the atom for the global control
 (something which is also referred to as {\em memoisation}).
In the case of the program $P$ from
 Example~\ref{example:parser} with ${\cal A}$ as depicted
 in Figure~\ref{figure:parser},
 we get the following program $P_{u}^{\cal A}$,
 where $(V_1,V_2)$ and $V_1$ represent $\sigma$ of
 Definition~\ref{def:pul-transf}:

{\small
\begin{verbatim}
  nont_u(X,T,R,(V1,V2)) :- t_u(a,T,V,V1),nont_m(X,V,R,V2).
  nont_u(X,T,R,V1) :- t_u(X,T,R,V1).
  t_u(X,[X|R],R,true).
\end{verbatim}
}
Suppose for the moment the simplest definition
 possible for {\tt\small nont\_m} 
 (i.e., it performs no global control nor does it filter and rename):
{\small
\begin{verbatim}
   nont_m(X,V,R,nont(X,V,R)).
\end{verbatim}
}

Evaluating the above code for the call
 {\tt\small nont\_u(c,T,R,Leaves)} then yields two computed answers which
 correspond to the two branches in Figure~\ref{figure:parser-unfold}
and allow us to reconstruct the unfiltered specialisation in
 Figure~\ref{figure:parser-spec}:

{\small
\begin{verbatim}
  > ?-nont_u(c,T,R,Leaves).
     T = [a|_A], Leaves = true,nont(c,_A,R) ? ;
     T = [c|R], Leaves = true ? ;
    no
\end{verbatim}
}


\subsection{The global control}
 \label{section:cogen:global}
 
As mentioned, the above code for $P_{u}^{\cal A}$
 is still incomplete, and we have to extend it to perform the global
control as well. Firstly, calling $p_u$ only returns the leaf atoms of one 
 branch of the SLDNF-tree, so we need to add some code that collects the
information from all the branches.%
This can be done very easily using Prolog's \svbt{findall} predicate.%
\footnote{Note that, because our generating extensions do not have to be
  self-applied, we do not necessarily
 have to specialise the \svbt{findall} predicate
  itself.}
 
In essence, \svbt{findall(V,Call,Res)} finds all the answers 
 $\theta_{i}$ of
 the call \svbt{Call}, applies $\theta_{i}$ to \svbt{V} and then
 instantiates \svbt{Res} to a list containing renamings of all the 
 \svbt{V}$\theta_{i}$'s.
In particular, \svbt{findall(B,nont\_u(c,T,R,B),Bs)}
 instantiates \svbt{Bs} to
 \svbt{[[true,nont(c,\_48,\_49)],} \svbt{[true]]}.
This essentially corresponds to the leaves of the SLDNF-tree in
 Figure~\ref{figure:parser-unfold} (by flattening and removing the $\mathit{true}$
 atoms we obtain
\svbt{[nont(c,\_48,\_49)]}). Furthermore, if we call
\svbt{findall(clause(nont(c,T,R),Bdy), nont\_u(c,T,R,Bdy), Cs)}
we will get in \svbt{Cs} a representation of the two resultants of
Figures~\ref{figure:parser-unfold} and
 \ref{figure:parser-spec} (without filtering).

Now, once all the resultants have been generated, the body atoms have to
 be generalised (using $\mathit{gen}_{\Delta}$) and then unfolded
 if they have not been encountered yet.
This is achieved by re-defining the predicates $p_m$
 so that they perform the global control.
That is, for every atom $p(\seq{t})$ in the original program, if one
 calls $p_m(\seq{t},R)$ then $R$ will be instantiated to the residual call
of $p(\seq{t})$ (i.e.\ the call after applying $\mathit{filter}_{\Delta}$;
 e.g., the residual call of $p(a,b,X)$ might be $p_1(X)$).
At the same time $p_m$ also generalises this call,
 checks if it has already been encountered, and if not, unfolds the atom
 to produce the corresponding residual code.
 
We have the following definition of $p_m$
 (we denote the Prolog conditional by $\mathit{If}\rightarrow \mathit{Then} ;  \mathit{Else}$):

\begin{definition} \label{def:pm-transf}
Let $P$ be a program and $p/n$ be a predicate defined in $P$.
Also, let $\seq{v}$ be a sequence of $n$ distinct variables (one for each 
argument of $p$).
We then define
  the clause $C^{p,\Delta}_{m}$ for $p_m$ as follows:

\vspace{1ex}
\renewcommand{\arraystretch}{\mikearraystretch}
\begin{tabular}{@{}l}
{\small\tt $p_m(\seq{v}$,R) :-}
 {\small\tt ( find\_pattern($p(\seq{v})$,R) -> {\tt true}}~\\
~~~~~~~~~~~~~~~~~{\small\tt ;~~(generalise($p(\seq{v})$,$p(\seq{g})$),}~\\
~~~~~~~~~~~~~~~~~~~~{\small\tt ~~insert\_pattern($p(\seq{g})$,Hd),}~\\
~~~~~~~~~~~~~~~~~~~~{\small\tt ~~ findall(clause(Hd,Bdy),$p_u(\seq{g}$,Bdy),Cs),}~\\
~~~~~~~~~~~~~~~~~~~~{\small\tt ~~ pp(Cs),}~\\
~~~~~~~~~~~~~~~~~~~~{\small\tt ~~ find\_pattern($p(\seq{v})$,R) ) ).}~
\end{tabular}
\vspace{1ex}

\noindent
Finally we define the Prolog program
  $P_{m}^{\Delta}$ =
 $\{ C^{p,\Delta}_{m} \mid p\in Pred(P) \}$.
\end{definition}

In the above, the predicate {\small\tt find\_pattern} checks whether its first
argument $p(\seq{v})$ 
 is an instance of a
 call that has already been specialised (or is in the process of being
 specialised) and, if it is,
its second argument
 will be instantiated to
 the properly renamed and filtered version
 $\mathit{filter}_{\Delta}(p(\seq{v}))$
 of the call.
This is the classical ``seen before'' check of partial evaluation
 \cite{Jones:peval}
 and is achieved by keeping a
list of the predicates that have been encountered before along with their renamed and
filtered calls. Thus, if the call to {\small\tt find\_pattern} succeeds,
 then {\small\tt R} has been instantiated to the
 residual call of $p(\seq{v})$, if the call was not
 seen before then the other branch of the
 conditional is executed.

The call {\small\tt generalise($p(\seq{v})$,$p(\seq{g})$)}
 simply computes $p(\seq{g}) = \mathit{gen}_{\Delta}(p(\seq{v}))$.

The predicate {\small\tt insert\_pattern} adds a new atom (its first
argument $p(\seq{g})$) to the list of atoms already encountered and returns (in its
second argument {\small\tt Hd}) the renamed and filtered version 
 $\mathit{filter}_{\Delta}(p(\seq{g}))$ of the generalised
 atom.
The atom {\small\tt Hd} will provide (maybe further instantiated)
 the head of the residual clauses.

This call to {\small\tt insert\_pattern}
 is put first to ensure that an atom is not specialised over and over again
 at the global level.
   
The call to
{\small\tt findall(clause(Hd,Bdy),$p_u(\seq{g}$,Bdy),Cs)}
 unfolds the generalised atom $p(\seq{g})$ and returns a list of
 residual clauses for $\mathit{filter}_{\Delta}(p(\seq{g}))$
 (in {\small\tt Cs}).
As we have seen in Section~\ref{section:cogen:local},
 the call to $p_u(\seq{g},Bdy)$
 inside this {\small\tt findall} returns one
 leaf goal
 of the SLDNF-tree for $p(\seq{g})$ at a time and instantiates
 $p(\seq{g})$ (and thus also {\tt\small Hd}) via the computed answer substitution of 
 the respective branch.
Observe that every atom $q(\overline{v})$
 in the leaf goal has already been renamed and filtered by 
 a call to the corresponding predicate $q_m(\overline{v})$.
 
Finally,
 the predicate {\small\tt pp} pretty-prints the clauses of the residual
program and the last call {\small\tt find\_pattern} will instantiate the output
 argument {\small\tt R} to the
 residual call $\mathit{filter}_{\Delta}(p(\seq{v}))$
 of the atom $p(\seq{v})$
(which is different from {\small\tt Hd} which is
 $\mathit{filter}_{\Delta}(p(\seq{g}))$).

We can now fully define what a generating extension is:

\begin{definition} \label{def:generating-extension}
  Let $P$ be a program and
  $(U_{{\cal A}},\Delta)$ a 
 strongly globally safe {\em BTC\/} for $P$, then the {\em
    generating extension\/} of $P$ with respect to
  $(U_{{\cal A}},\Delta)$ is the Prolog program
  $P_{g} = P_{u}^{\cal A} \cup P_{m}^{\Delta}$.
\end{definition}

\noindent The generating extension is called as follows: if one wants to
specialise an atom $p(\seq{v})$ one simply calls $p_m(\seq{v},${\tt\small R}).
Observe that generalisation and specialisation occur
 {\em as soon as\/} we call $p_m(\seq{v},${\tt\small R}), and not
 after the whole incomplete SLDNF-tree has been built.%
\footnote{It is, however, not very difficult to change 
 the $\mathit{cogen}$ so that it calls $p_m(\seq{v},${\tt\small R})
 only after the whole incomplete SLDNF-tree has been built.}
Together with our particular construction of the unfolder predicates 
 (Definition~\ref{def:pul-transf}) this means that to ensure 
 correctness of specialisation we need to have strong
 global safety instead of just
 global safety (cf.\ Definitions~\ref{def:safety} and \ref{def:strong-global-safety}).

There are several ways to improve the definition
 of a generating
 extension (Definition~\ref{def:generating-extension}).
The first improvement relates to the call
 {\small\tt generalise($p(\seq{v})$,$p(\seq{g})$)}
  which computes $p(\seq{g})$ = $\mathit{gen}_{\Delta}(p(\seq{v}))$.
If the division for $p$ 
in $\Delta$ is 
 simple (i.e., only contains {\tt\small static} and 
{\tt\small dynamic}) one can actually compute $p(\seq{g}) =
\mathit{gen}_{\Delta}(p(\seq{v}))$
 beforehand (i.e., in the $\mathit{cogen}$ as opposed to in the 
 generating extension), without having to know the actual values for the 
 variables in $\seq{v}$.
This will actually be used by our $\mathit{cogen}$,
 whenever possible, to further improve the efficiency of the generating 
 extensions.
For example, if we have $\Delta = \{p(\mathit{static},\mathit{dynamic})\}$ and
 $p(\seq{v})$ = {\small\tt p(X,Y)}, then the
$\mathit{cogen}$ does not have to generate a call to {\tt\small generalise/2};
it can simply use {\small\tt p(X,Z)} for $p(\seq{g})$,  where Z is a fresh variable,
within the code for {\small\tt p\_m(X,Y,R)}.
The generating extension will thus correctly keep the static values in 
{\small\tt X} and abstract
 the dynamic values in {\small\tt Y}.

Second, in practice it might be unnecessary to define 
  $p_{m}$ for every predicate $p$.
Indeed, there might be predicates which are never memoised.
Such predicates will never appear at the global level, and one can 
 safely remove the corresponding definitions for $p_{m}$ from
  Definition~\ref{def:generating-extension}.

For instance, in Example~\ref{example:parser} the predicate $t/3$ is 
always reducible and never specialised immediately by the user.
Also, the division is 
 simple, and one can thus 
pre-compute {\small\tt generalise}.
The resulting, optimised generating extension is
 shown in Figure~\ref{figure:parser-generating-extension}.

\begin{figure}[thb]
\begin{center}
\begin{small}
\begin{verbatim}
    nont_m(B,C,D,FilteredCall) :- 
       (find_pattern(nont(B,C,D),FilteredCall) -> true
        ; (insert_pattern(nont(B,F,G),FilteredHead),
           findall(clause(FilteredHead,Body),
                   nont_u(B,F,G,Body),SpecClauses),
           pp(SpecClauses),
           find_pattern(nont(B,C,D),FilteredCall)
       )).
    nont_u(B,C,D,(E,F)) :- t_u(a,C,G,E),nont_m(B,G,D,F).
    nont_u(H,I,J,K) :- t_u(H,I,J,K).
    t_u(L,[L|M],M,true).
\end{verbatim}
\end{small}
\caption{The generating extension for the parser
\label{figure:parser-generating-extension}}
\end{center}
\end{figure}

\subsection{The $\mathit{cogen}$ {\sc logen}}

The job of the $\mathit{cogen}$
is now quite simple: given a program $P$ and a strongly globally safe
{\em BTC\/} $\beta$ for $P$, produce a generating extension for $P$ consisting
of the two parts described above. The code of the essential
parts of our $\mathit{cogen}$, called
 {\sc logen},
 is shown in Appendix~\ref{appendix:cogen}.  The
predicate {\tt\small memo\_clause} generates the definition of the global control
$m$-predicates for each non-reducible predicate of the program whereas the
predicates {\tt\small unfold\_clause} and {\tt\small body} take care of translating
clauses of the original predicate into clauses of the local control
$u$-predicates. Note how the second argument of {\tt\small body}
corresponds to code of the generating extension whereas the third argument
corresponds to code produced at the next level, i.e.\ at the level of the
specialised program.

\subsection{An Example}

We now show that {\sc logen} is actually powerful 
enough to satisfactorily specialise the vanilla metainterpreter
 (a task which has attracted a lot of attention
\cite{BUP-Meta-Pe:PDK91,MartensDeSchreye:jlp95b,VanHoofMartens:Parse}
 and is far from trivial).
 
\begin{example}
    The following is the well-known vanilla metainterpreter for the 
    non-ground representation, along with an encoding of the
 ``double append'' program:
    
\begin{small}\begin{tt}
\begin{pitemize}
\item demo(true).
\item demo((P \& Q)) :- demo(P), \underline{demo}(Q).
\item demo(A) :- \underline{dclause}(A,Body), \underline{demo}(Body).
\item  ~
\item dclause(append([],L,L),true).
\item dclause(append([H|X],Y,[H|Z]),append(X,Y,Z) \& true).
\item dclause(dapp(X,Y,Z,R), (append(X,Y,I) \& (append(I,Z,R) \& true))).
\end{pitemize}
\end{tt}\end{small}

Note that in a setting with just the static/dynamic binding types
 one cannot
specialise this program in an interesting way, because the argument to
{\tt\small demo} may (and usually will) contain variables.
This is why neither \cite{JorgensenLeuschel:Cogen} nor
 \cite{MogensenBondorf:LOPSTR92} were able to handle this example.
We, however, can produce the {\em BTC\/} $({\cal A},\Delta)$ with
$\Delta = \{\mathit{demo}(\mathit{nonvar}),$
 $\mathit{dclause}(\mathit{nonvar},$ $\mathit{dynamic})\}$ and where
the  annotation
${\cal A}$ is such that every literal
 but the $\mathit{demo}(P)$ call in the second clause is marked as reducible
(see underlining above).

Observe that, to make the {\em BTA\/} simpler, we  encode conjunctions
  in a  list-like fashion
  within the second argument of $\mathit{dclause}$ as follows:
 a conjunction $A_1 \wedge \ldots \wedge A_n$ will be
 represented as
 $A_1 \& ( \ldots (A_n \& \mathit{true}))$.
This enables us to separate the conjunction skeleton from the
 individual literals, and allows us to produce an annotation which
 will result in removing all the parsing overhead related to the
 conjunction skeleton but will not unfold potentially recursive literals
 within the conjunctions.

The importance of the $\mathit{nonvar}$ annotation is its influence on
the generalisation operation.
Indeed,
 we have
 $\mathit{gen}_{\Delta}(\mathit{demo}(\mathit{append}(X,[a],Z)))$
 = $\mathit{demo}(\mathit{append}(X,Y,Z))$
 whereas for
 $\Delta' = \{\mathit{demo}(\mathit{dynamic}),$
 $dclause(\mathit{dynamic},$ $\mathit{dynamic})\}$ the generalisation
 operation throws away too much information:
$\mathit{gen}_{\Delta'}(\mathit{demo}(\mathit{append}(X,[a],Z)))$ 
 = $\mathit{demo}(C)$, resulting in very little specialisation.

The {\tt\small demo\_u} unfolder predicate generated by the $\mathit{cogen}$
 for {\tt\small demo} then looks like:

\begin{small}\begin{tt}
\begin{pitemize}
\item demo\_u(true,true).
\item demo\_u(B \& C,(D,E)) :- demo\_m(B,D), demo\_u(C,E).
\item demo\_u(F,(G,H)) :- dclause\_u(F,I,G), demo\_u(I,H).
\end{pitemize}
\end{tt}\end{small}

The specialised code that is produced by the generating extension
 (after flattening) for 
 the call $\mathit{demo}(\mathit{dapp}(X,Y,Z,R))$ is:

\begin{small}\begin{tt}
\begin{pitemize}
\item demo\_\_0(B,C,D,E) :- demo\_\_1(B,C,F), demo\_\_1(F,D,E).
\item demo\_\_1([],B,B).
\item demo\_\_1([C|D],E,[C|F]) :-  demo\_\_1(D,E,F).
\end{pitemize}
\end{tt}\end{small}

Observe that specialisation has been successful:
 all the overhead has been compiled away and {\small\tt demo\_\_1} 
 even corresponds to the definition of {\tt\small append}.
Given the above {\em BTC\/}, {\sc logen} can achieve a similar 
 feat for {\em any\/} object program and query to 
 be specialised.
 As we will see in Section~\ref{section:benchmarks}
  it can do so efficiently.

Finally, note that the inefficiency of traversing the first
 argument to $\mathit{dapp}$ twice has not been removed.
 For this, conjunctive partial deduction is needed
 \cite{CPD:megapaper}.

\end{example}


\section{Extending {\sc logen}}
\label{section:cogen-impure}

In this section we will describe how to extend {\sc logen} 
to handle logic programming languages with 
  built-ins and non-declarative features. We will
explain these extensions for Prolog, but many of the ideas should also
carry over to other logic programming languages.
(Proponents of Mercury and G\"{o}del may safely skip all
 but Subsection~\ref{section:declarative-primitives}.)

\subsection{Declarative primitives}
\label{section:declarative-primitives}

It is straightforward to extend {\sc logen}
 to handle declarative primitives,
i.e.\ built-ins such as {\tt\small =/2}, {\tt\small is/2} and {\tt\small arg/3},%
\footnote{E.g., {\tt arg/3} can be viewed as being defined by 
 a (possibly infinite) series of facts:
 {\tt arg(1,h(X),X).}, {\tt arg(1,f(X,Y),X).}, {\tt arg(2,f(X,Y),Y).},
  \ldots}
 or externally defined
 user predicates (i.e., predicates defined in another file or
 module,%
\footnote{Of course, doing a modular binding-time analysis
 is more difficult than doing an ordinary one,
 but it is possible \cite{Vanhoof:LPAR2000} and this is not
 really our concern here.} as
long as these are declarative).

The code of these predicates is not
 available to the $\mathit{cogen}$ and therefore no predicates
to unfold them can be
generated. The generating extension can therefore do one of two things:
\begin{enumerate}
    \item either
 completely evaluate a call to such primitives 
 (reducible case), 
 \item or simply produce a residual call
  (non-reducible case). 
\end{enumerate}
To achieve this, we simply extend
 the transformation of Definition~\ref{def:pul-transf} with
 the following two rules, where $c$ is a call to a declarative 
 primitive and reducible calls are underlined:

$$
\begin{array}{ll}
    \underline{c} \specialise c : \mathit{true} \\[1ex]
    c \specialise \mathit{true} : c
\end{array}
$$

\begin{example}
For instance, we have
 $\underline{{\tt\small arg(1,X,A)}} \specialise {\tt\small arg(1,X,A)} : \mathit{true}$,
 meaning that the call will be executed in the generating extension 
 and nothing has to be done in the specialised program.
On the other hand, we have
 ${\tt\small arg(N,X,A)} \specialise \mathit{true} : {\tt\small arg(N,X,A)}$,
 meaning that the 
 call is only executed within the specialised program.
Now take the clause:\\
\hspace*{0.5cm} {\tt\small p(X,N,A) :- \underline{arg(1,X,A)},arg(N,X,A)}.\\
This clause is transformed (by~ $\specialise_{u}$) into the following 
unfolding clause:\\
\hspace*{0.5cm} {\tt\small p\_u(X,N,A,arg(N,X,A)):- arg(1,X,A).}\\
For $\Delta$ = $\{p(\mathit{static},\mathit{dynamic},\mathit{dynamic}\}$ and
 for {\tt\small X = f(a,b)} the generating extension will produce the residual 
code:\\
\hspace*{0.5cm}{\tt\small p\_\_0(N,a) :- arg(N,f(a,b),a).}\\
while for $X = a$ the call {\tt\small arg(1,a,A)} will fail
 and no code will be 
produced (i.e., failure has already been detected within the 
generating extension).

Observe that, while {\tt\small arg(1,a,A)} fails in
 SICStus Prolog, it actually raises an error in ISO Prolog.
So, in the latter case we actually have to generate a
 residual clause of the form {\tt p\_\_0(N,A) :- raise\_exception(...).}
\end{example}

\subsection{Problems with non-declarative primitives}

The above two rules could also be used for non-declarative primitives.
However, the code generated will in general be 
incorrect, for the following two reasons.

First, for some calls $c$ to non-declarative primitives
 $c,\mathit{fail}$ is not equivalent to $\mathit{fail}$.
For example, 
 {\tt\small print(a),fail} behaves differently from {\tt\small fail}.
Predicates $p$ for which the conjunctions
 $p(\seq{t}),\mathit{fail}$ and $\mathit{fail}$ are not equivalent
  are termed as ``side-effect'' in \cite{Sahlin93:ngc}.
For such predicates the 
independence on the computation rule does not hold.
In the context of the Prolog left-to-right computation 
 rule, this means that we have to ensure 
 that failure to the right of such a call
 $c$ does not prevent the generation of the 
 residual code for $c$ nor its execution at runtime.
For example, the clause \\
\hspace*{0.5cm} {\tt\small t :- print(a), \underline{2=3}.}\\
can be specialised to
 ~{\tt\small t :- print(a),fail.} but not to
 ~{\tt\small t :- fail,print(a).}
 and neither to
 ~{\tt\small t :- fail.} nor to the empty 
 program.
The scheme of Section~\ref{section:declarative-primitives} 
 would produce the following
 unfolder predicate, which is incorrect
  as it produces the empty program:\\
\hspace*{0.5cm} {\tt\small t\_u(print(a)) :- 2=3.}

The second problem are the so called ``propagation sensitive'' 
\cite{Sahlin93:ngc} built-ins.
For calls $c$ to such built-ins, even though
 $c,\mathit{fail}$ and $\mathit{fail}$ are equivalent, the
 conjunctions
 $c, ~ X=t$ and $X=t, ~ c$ are not.
One such built-in is {\tt\small var/1}:
 we have, e.g., that
 {\tt\small (var(X),X=a)} is not equivalent to {\tt\small (X=a,var(X))}.
Again,
 independence on the computation rule is violated 
 (even though there are no side-effects), which again
  poses problems for specialisation.
Take for example the following clause:\\
\hspace*{0.5cm} {\tt\small t(X) :- var(X), \underline{X=a}.}\\
The scheme of Section~\ref{section:declarative-primitives} 
 would produce the following
 unfolder predicate:\\
\hspace*{0.5cm} {\tt\small t\_u(X,var(X)) :- X=a.}\\
Running this for {\tt\small X} uninstantiated will produce the 
 following residual
 code, which is incorrect as it always fails:\\
\hspace*{0.5cm} {\tt\small t(a) :- var(a).}\\
To solve this problem we will have to ensure that
 bindings generated by specialising
 calls to the right of propagation sensitive 
 calls $c$ do not backpropagate
 \cite{Sahlin93:ngc,Prestwich92:TR} onto $c$.
In the case above, we have to prevent the binding {\tt\small X/a} to 
backpropagate onto the {\tt\small var(X)} call.

In the remainder of this section we show how side-effect and 
 propagation sensitive predicates can 
  be dealt with in a rather elegant and still efficient manner in 
 our $\mathit{cogen}$ approach.

\subsection{Hiding failure and sensitive bindings} 

To see how we can solve our problems, we examine a small example 
in more detail. Take the following program:\\
\hspace*{0.5cm} {\tt\small p(X) :- {print(X)},{var(X)}, 
\underline{q(X)}.}\\
\hspace*{0.5cm} {\tt\small q(a)}.\\
We have that \underline{{\tt\small q(X)}}
 $\specialise$ {\tt\small q\_{u}(X,C)} : {\tt\small C},
 and applying the scheme from Section~\ref{section:declarative-primitives}
 naively, we get:\\
\hspace*{0.5cm}{\tt\small p\_{u}(X,(print(X),var(X),C)) :- q\_u(X,C).}\\
\hspace*{0.5cm}{\tt\small q\_{u}(a,true).}\\
For the same reasons as in the above examples this unfolder predicate 
is incorrect (e.g., for {\tt\small X=b} the empty program is generated).

To solve the problem we have to 
 avoid
 backpropagating 
 the bindings generated by
 {\tt\small q\_u(X,C)} onto {\tt\small print(X),var(X)}
 and ensure that a failure of
 {\tt\small q\_u(X,C)} does not prevent code being generated for {\tt\small print(X)}.
The solution is to wrap {\tt\small q\_u(X,C)} into a
 call to {\tt\small findall}.
Such a call will not instantiate {\tt\small q\_u(X,C)}
 and if {\tt\small q\_u(X,C)} fails this will only lead to the third argument of
 {\tt\small findall} being instantiated to an empty list.
\ignore{****
A first attempt might thus look like this:\\
\hspace*{0.5cm}{\tt\small p\_{u}(X,(print(X),var(X),Cs)) :- }
   {\tt\small findall(C,q\_u(X,C),Cs).}\\
\hspace*{0.5cm}{\tt\small q\_{u}(a,true).}\\
If we now run {\tt\small p\_u(X,Code)}
we get the residual code:\\
\hspace*{0.5cm} {\tt\small p(X) :- {print(X)},{var(X)},{true}.}\\
So, backpropagations have been prevented but unfortunately there is 
also now no link at all between the instantiations performed by
{\tt\small q\_u(X,C)} and the rest of the clause.

To arrive at a full solution we have to re-create this link
 within the residual code (but {\em not\/} within the generating 
 extension).
If we knew that {\tt\small q\_u(X,C)} would always have exactly one solution
 (which it has not; it can fail) we could write the following 
 code:\\
\hspace*{0.5cm}{\tt\small p\_{u}(X,(print(X),var(X),X=Xs,Cs)) :-}\\
\hspace*{1.5cm} {\tt\small findall((X,C),q\_u(X,C),[(Xs,Cs)]).}\\
\hspace*{0.5cm}{\tt\small q\_{u}(a,true).}\\
Notice how we have added the variables of {\tt\small q(X)}
 as an extra argument to the {\tt\small findall} and how
  {\tt\small X=Xs} re-creates the link between the variables in {\tt\small Cs}
  and the rest of the clause.
Running {\tt\small p\_u(X,Code)} produces the correct residual code:\\
\hspace*{0.5cm} {\tt\small p(X) :- {print(X)},{var(X)},{X=a,true}.}
***}
To link up the solutions of the {\tt\small findall} with the rest
of the unfolding process we use an auxiliary predicate
 {\tt\small make\_disjunction}.
All this leads to the following extra rule, to be added
 to Definition~\ref{def:pul-transf}, and where calls whose bindings 
 and whose failure should be hidden are wrapped into a
  {\tt\small hide\_nf} annotation:

$$
\begin{array}{ll}
    \begin{array}{c}
        \kappa \specialise \gamma : \sigma\\
        \hline
       hide\mbox{{\tt \_}}nf(\kappa) \specialise \\
       \mbox{{\tt\small varlist}}(\kappa,V),\\
       \mbox{{\tt\small findall}}((\sigma,V),\gamma,R),\\
       \mbox{{\tt\small make\_disjunction}}(R,V,C)\\
       : C 
    \end{array} & R,V,C \mbox{~fresh variables}
\end{array}
$$

The full code of {\tt\small make\_disjunction} is straightforward and
 can be found in Appendix~\ref{appendix:cogen}.
 
One might wonder why in the above solution one just
 keeps track of the variables in $\kappa$.
The reason is that
 all the variables in $\gamma$ or $\sigma$  (in contrast to $\kappa$)
 cannot occur in the 
 remainder of the clause.

Note that annotating a call $c$ using
 {\tt\small hide\_nf} also prevents right-propagation of 
 bindings generated while specialising $c$.
This is not a restriction, because instead of writing
 {\tt\small hide\_nf}($\alpha$),$\beta$
 we can always write
 {\tt\small hide\_nf}(($\alpha$,$\beta$))
 if one wants the instantiations of $\alpha$ to be propagated 
 onto $\beta$.
Furthermore, preventing right-propagations
 will turn out to be useful in the treatment of negations,
  conditionals, and disjunctions below.

\begin{example} 
    Let us trace the thus extended $\mathit{cogen}$ on another example:

\begin{small}\begin{verbatim}
  p(X) :- print(X), q(X).
  q(a).
  q(b).
\end{verbatim}\end{small}

Let us mark {\tt\small q(X)} as reducible and wrap it
 into a {\tt\small hide\_nf()} annotation; 
 the exact representation of the annotated clause
  required for {\sc logen} is:

\begin{small}\begin{verbatim}
 ann_clause(1,p(X),(rescall(print(X)),hide_nf(unfold(q(X))))).
\end{verbatim}\end{small}

\noindent
We now get the following unfolding predicate
for {\tt\small p}:

\begin{small}\begin{verbatim}
 p_u(X,(print(X),Disj)) :- 
     varlist(q(X),Vars),
     findall((Code,Vars), q_u(X,Code), Cs),
     make_disjunction(Cs,Vars,Disj).
\end{verbatim}\end{small}

If we run the generating extension we get the residual program
 (calls to {\tt\small true} have been removed by the $\mathit{cogen}$):

\begin{small}\begin{verbatim}
  p__0(B) :- print(B), (B = a ; B = b).
\end{verbatim}\end{small}

\end{example}

Instead of generating disjunctions, one could also produce
 new predicates for each disjunction (at least for those
 cases where argument indexing might be lost
 \cite{VenkenDemoen:ngc88}).

\subsection{A solution for non-leftmost, non-determinate
 unfolding}
\label{section:expensive_pred}

It is well known that non-leftmost, non-determinate unfolding,
 while sometimes essential for satisfactory propagation of
 static information, can cause substantial slowdowns.
Below we show how our new {\tt\small hide\_nf} annotation
 can solve this dilemma (another solution is conjunctive
 partial deduction \cite{LeuschelDeSchreyeDeWaal:JICSLP96}).

\begin{example} 
   In the following {\tt\small expensive\_predicate(X)}
    is an expensive, but fully declarative predicate,
     which for some reason (e.g., termination) we cannot unfold.

\begin{small}\begin{verbatim}
  p(X) :- expensive_predicate(X), q(X), r(X).
  q(a).          r(a).
  q(b).          r(b).
  q(c).
\end{verbatim}\end{small}

If we mark {\tt\small expensive\_predicate(X)} as non-reducible,
 and {\tt\small q(X)} and {\tt\small r(X)}  as reducible we get the following
residual
  program:

\begin{small}\begin{verbatim}
  p__0(a) :- expensive_predicate(a).
  p__0(b) :- expensive_predicate(b).
\end{verbatim}\end{small}

This residual program has left-propagated the bindings,
 which is not a problem in itself, but potentially duplicates
  computations and leads to a less efficient residual program.
A solution to this problem, which still allows
 one to unfold {\tt\small q(X)} (and right-propagate the bindings onto
 {\tt\small r(X)}) and {\tt\small r(X)} is to
wrap them
 into a {\tt\small hide\_nf} annotation.
This is represented as the following annotated clause,
 where {\tt unfold} is wrapped around calls to be unfolded
 and {\tt rescall} is wrapped around non-reducible primitives:

\begin{small}
\begin{verbatim}
  ann_clause(1,p(X),(rescall(expensive_predicate(X)),
                      hide_nf((unfold(q(X)),unfold(r(X)))))).
\end{verbatim}
\end{small}

We then get the following residual program:

\begin{small}\begin{verbatim}
  p__0(B) :- expensive_predicate(B), (B = a ; B = b).
\end{verbatim}\end{small}
\end{example}

\subsection{Generating correct annotations}
\label{section:generating_anns}

Having solved the problem of left-propagation of failure and bindings,
 we now just have to figure out when {\tt\small hide\_nf} annotations
 are actually necessary.
In order to achieve maximum specialisation and efficiency, one would 
want to use just the minimum number of such annotations which still 
ensures correctness.

First, we have to
 define a new relation $\models_{\mathit{hide}} \gamma$ that
holds if the code $\gamma$ within the generating extension cannot
fail and cannot instantiate variables in the remainder of the generating
extension. This relation is defined in
 Figure~\ref{figure:nf-relation}.
This definition can actually be kept quite simple
 because it is intended to be applied to code in
 the generating extension which has a very special form. 

The following modified rule for conjunctions
 (replacing the corresponding rule in
  Definition~\ref{def:pul-transf})
  ensures that no 
 bindings are left-propagated or side-effects removed.
$$
\begin{array}{ll}
    \begin{array}{c}
      \kappa_{i} \specialise \gamma_{i} : \sigma_{i} 
      ~\wedge~
      \mathit{impure}(\kappa_{i}) \Rightarrow \forall j>i: \models_{\mathit{hide}}\gamma_{j}
      \\
      \hline
      (\kappa_{1}, \ldots , \kappa_{n}) \specialise 
      (\gamma_{1}, \ldots , \gamma_{n}) : (\sigma_{1}, \ldots , 
      \sigma_{n})
    \end{array} 
    \\[2ex]
\end{array}
$$
Here $\mathit{impure}(\kappa_{i})$ holds if $\kappa_{i}$ contains a call to a
side-effect predicate (which has to be non-reducible) or to a 
 non-reducible propagation sensitive call.
Calls are classified as in \cite{Sahlin93:ngc} (e.g., the property of 
 generating a side-effect propagates up the dependency graph).
In case we want to prevent backpropagation of bindings on expensive
 predicates as discussed in Section~\ref{section:expensive_pred},
 then $\mathit{impure}(\kappa_{i})$ should also hold when
 $\kappa_{i}$ contains a call to a non-reducible, expensive predicate.

This modified rule for conjunctions together with
 Figure~\ref{figure:nf-relation} can be used to determine the required
 {\tt\small hide\_nf} annotations.
For example, the first rule in Figure~\ref{figure:nf-relation}
 actually implies that non-reducible calls never pose a problem
 and do not have to be wrapped into
 a {\tt\small hide\_nf} annotation (because they produce
 the code $\gamma_i$ = $\mathit{true}$ within the generating extension).

\begin{figure}
\begin{center}
$$
\begin{array}{ll}
    \begin{array}{c}
       ~\\
              \hline
              \models_{\mathit{hide}} \mathit{true} 
    \end{array}

    \\[8ex]

    \begin{array}{c}
       \forall i: ~ \models_{\mathit{hide}} \gamma_{i}\\
              \hline
              \models_{\mathit{hide}} \gamma_{1},\ldots,\gamma_{n} 
    \end{array} 
~~~
 &
~~~

    \begin{array}{c}
      \forall i: ~ \models_{\mathit{hide}} \gamma_{i}\\
              \hline
              \models_{\mathit{hide}} ( \gamma_{1} ; \gamma_{2}) 
    \end{array} 

    \\[8ex]
    \begin{array}{c}
      \forall i: ~ \models_{\mathit{hide}} \gamma_{i}\\
              \hline
              \models_{\mathit{hide}} (\gamma_{1} \rightarrow \gamma_{2} ; \gamma_{3}) 
    \end{array} 
~~~
&
~~~
    \begin{array}{c}
       hide\mbox{{\tt \_}}nf(\kappa) \specialise \gamma : \sigma\\
              \hline
              \models_{\mathit{hide}} \gamma 
    \end{array} 
\end{array}
$$
\caption{The hide relation
$\models_{\mathit{hide}}$\label{figure:nf-relation}\label{figure:hide-relation}}
\end{center}
\end{figure}

To further improve specialisation and efficiency
 one could also introduce additional annotations such as
  $\texttt{nf}(\kappa)$
  if only non-failing has to be prevented  and
   $\texttt{hide}(\kappa)$
  if only bindings have to be hidden.
This is actually done within the implementation of the $\mathit{cogen}$, but, 
for clarity's sake, we don't elaborate on this here.

\subsection{Negation and Conditionals}

Prolog's negation (not/1) is handled similarly to a declarative primitive,
except that for the residual case $not(\kappa)$ we will also specialise the
code $\kappa$ inside the negation and we have to make sure that this
specialisation (performed by the generating extension) cannot
fail (otherwise the code generation would be incorrectly prevented)
or propagate bindings.

$$
\begin{array}{ll}
   \begin{array}{c} \kappa \specialise \gamma : \mathit{true} \\ \hline
        \underline{\texttt{not}}(\kappa) \specialise \texttt{not}(\gamma) :
        \mathit{true} \end{array} 
\\[7ex]
 \begin{array}{c} \kappa \specialise \gamma
        : \sigma ~~\wedge~~ \models_{\mathit{hide}} \gamma\\ \hline
        \texttt{not}(\kappa) \specialise \gamma : \texttt{not}(\sigma)
        \end{array}
\end{array}
$$

The first rule is used when we know that $\kappa$ can be completely
 and finitely unfolded and it can
 be determined whether $\kappa$ fails or not:
 if $\gamma$ succeeds then the generating
 extension will not generate code, and if $\gamma$ fails the
 generating extension will succeed and produce the residual code 
 {\tt\small true} for the negation.
If we have $\kappa \specialise \gamma : \sigma$ with $\sigma\neq\mathit{true}$
 then the annotation was wrong and an error will be raised
 during specialisation.
It is thus the responsibility of the BTC to ensure that such errors do not
occur.

If the negation is non-reducible then we require that the generating
extension does not fail (the hide relation in the premiss). To enable
the rule, $\kappa$ must be given the {\tt\small hide\_nf} annotation
unless $\gamma$ is already hidden.
Again, this is the responsibility of the BTC.

\begin{example}
Consider the following two annotated clauses.
$$
\begin{array}{lcl}
\mbox{\tt\small p(X)} & \mbox{\tt\small :-} &
 \underline{\mbox{\tt\small not}}\mbox{\tt\small (X}\underline{\mbox{\tt\small
=}}\mbox{\tt\small a)}.\\
\mbox{\tt\small q(Y)} & \mbox{\tt\small :-} & \mbox{\tt\small not(Y=a)}.
\end{array}
$$
In the first clause {\tt\small X} is assumed to be of binding-type {\tt\small static}
 (or at least {\tt\small nonvar}) so the
 negation can be reduced.%
\footnote{Note that it is up to the binding-type analysis to mark negations as 
reducible only if this is sound, e.g., when the arguments are ground.}
In the second we assume that {\tt\small Y} is
 dynamic.
If we run the generating extension with goal {\tt\small p({\tt\small a})} we will
get an empty program, which is correct.
If we run the generating extension with goal {\tt\small q(Y)} we will
get the following (correct) residual clause: 
\begin{verbatim}
  q__0(B) :- not(B=a).
\end{verbatim}
\end{example}
 
Handling conditionals is also straightforward.
If the test goal of a conditional is reducible then we can evaluate the
 conditional within the generating extension.
If the test goal of the conditional
 is non-reducible then, similarly to the negation, we require that the
three subgoals in the generating extension do not fail nor propagate 
bindings:
$$
\begin{array}{ll}
    \begin{array}{c}
      \forall i: ~ \kappa_{i} \specialise \gamma_{i} : \sigma_{i} \\
      \hline
      (\kappa_{1} \underline{\texttt{->}} \kappa_{2}
                  \underline{\texttt{;}} \kappa_{3})
       \specialise 
      (\gamma_{1} \texttt{->} (\gamma_{2}, \sigma_{2} \texttt{=} C) \texttt{;}
                              (\gamma_{3}, \sigma_{3} \texttt{=} 
                              C)) : C
    \end{array}
    & \mbox{(C fresh variable)}
    \\[7ex]
    \begin{array}{c}
      \forall i: ~ \kappa_{i} \specialise \gamma_{i} : \sigma_{i} ~~\wedge~~
       \models_{\mathit{hide}} \gamma_{i}\\
      \hline
      (\kappa_{1} \texttt{->} \kappa_{2} \texttt{;}
                                          \kappa_{3}) \specialise 
      \gamma_{1},\gamma_{2},\gamma_{3} : (\sigma_{1} \texttt{->} \sigma_{2} \texttt{;} \sigma_{3})
    \end{array}
    \\[2ex]
\end{array}
$$

\subsection{Disjunctions}
\label{subsection:disjunction}



To handle disjunctions we will use  our 
 {\tt\small hide\_nf} annotation to ensure that failure
 of one disjunct does not cause the whole specialisation to fail.
It will also ensure that the bindings from one disjunct do not propagate
 over to other disjuncts.
The rule for disjunctions therefore has the form:
$$
\begin{array}{ll}
    \begin{array}{c}
     \forall i: ~ \kappa_{i} \specialise \gamma_{i} : \sigma_{i} ~~~~ 
      \models_{\mathit{hide}} \gamma_{i}\\
      \hline
      (\kappa_{1}; \ldots ; \kappa_{n}) \specialise 
      (\gamma_{1}, \ldots , \gamma_{n}) : (\sigma_{1}; \ldots ; 
      \sigma_{n})
    \end{array}
\end{array}
$$

The above rule will result in a disjunction
being created in the residual code.
We could say that the disjunctions are
residualised. It is possible to treat disjunction in a different way in
which they are reduced away, but at the price of some
 duplication of work and residual code.
The rule for such {reducible disjunctions} is:
$$
\begin{array}{ll}
    \begin{array}{c}
     \forall i: ~ \kappa_i \specialise \gamma_i : \sigma_i\\
      \hline
      (\kappa_1\underline{;} \ldots \underline{;} \kappa_n) \specialise 
      (\gamma_1, \sigma_1\mbox{\tt\small =C}; \ldots ; \gamma_n, \sigma_n\mbox{\tt\small
=C}) : 
      \mbox{\tt\small C}
    \end{array} & \mbox{({\tt\small C} fresh variable)}
\end{array}
$$
The drawback of this rule is that it may duplicate work and code. To
see this consider a goal of the form:
$Q_h,(Q_1;Q_2),Q_t$ .
If specialisation of $Q_1;Q_2$ does not give any instantiation of the
variables that occur in $Q_h$ and $Q_t$ then these will be specialised
twice and identical residual code will be generated each time.

\subsection{More refined treatment of the {\tt call} predicate}
\label{section:call-pred}

In this section we present one example of specialisation using the
 {\tt\small call} predicate and show how its specialisation can be
 further improved.
The {\tt\small call} predicate can be considered to be declarative%
\footnote{If delayed until its argument is {\tt\small nonvar},
 it can be viewed as being defined by 
 a series of facts:
 {\tt\small call(p(X)) :- p(X).}, {\tt\small call(q(X,Y)) :- q(X,Y).},
  \ldots}
 and is important for implementing higher-order primitives in Prolog.
Unfortunately, current implementations of {\tt\small call} are not very 
efficient and it would therefore be ideal if the overhead could be 
removed by specialisation.
This is exactly what we are going to do in this section.
 
In {\tt\small call(C)} the value of {\tt\small C} can
either be a call to a built-in or a user-defined predicate. Unless the
predicate is externally defined the two cases require different
treatment. 
Consider the following example, featuring
 the Prolog implementation of the higher-order map predicate:

{\small
\begin{verbatim}
  map(P,[],[]).
  map(P,[H|T],[PH|PT]) :- Call =.. [P,H,PH], call(Call), map(P,T,PT).
  inc(X,Y) :- Y is X + 1.
\end{verbatim}}

Assume that we want to specialise the call {\tt\small map(inc,I,O)}. We can
produce the BTC $({\cal A},\Delta)$ with $\Delta = \{\mbox{\tt\small
map}(\mathit{static},\mathit{dynamic},\mathit{dynamic}), \mbox{\tt\small
inc}(\mathit{dynamic},\mathit{dynamic})\}$ and where ${\cal
A}$ marks everything, but the {\tt\small =../2} call in clause 2, as non-reducible.
Indeed,
 since the value of {\tt\small Call} is not known when we generate the unfolding
predicate for {\tt\small map} we should in general not try to unfold the atom bound to
{\tt\small Call}. The unfolding predicate generated by the $\mathit{cogen}$
  thus looks like:

{\small
\begin{verbatim}
  map_u(B,[],[],true).
  map_u(C,[D|E],[F|G],(call(H),I)) :- H =.. [C,D,F], map_m(C,E,G,I).
\end{verbatim}}

The specialised code obtained for the call {\tt\small map(inc,I,O)} is:

{\small
\begin{verbatim}
  map__0([],[]).
  map__0([B|C],[D|E]) :- inc(B,D), map__0(C,E).
\end{verbatim}}

All the overhead of {\tt\small call} and {\tt\small =..}
 has been specialised away, but one still needs the
original program to evaluate {\tt\small inc}.
To overcome this limitation, one can devise a special treatment for 
 calls to user-defined predicates
 which enables unfolding {\em within\/} a  {\tt\small call/1} primitive:
$$
\begin{array}{ll}
    \underline{call}(A) \specialise
     \mbox{\tt\small add\_extra\_argument}("_u",A,C,G),call(G) : C & 
    \mbox{(C fresh variable)}\\
    call(A) \specialise \mbox{\tt\small add\_extra\_argument}("_m",A,C,G),call(G) : C & 
    \mbox{(C fresh variable)}
\end{array}
$$
In both cases the argument to {\tt\small call} has to be a 
user-defined predicate which will be known by the
generating extension but is not yet known at $\mathit{cogen}$ time.
If this is not the case one has to use
 the standard technique for built-ins and possibly keep the
 original program at hand.

The code for  {\tt\small add\_extra\_argument} can be found in
 Appendix~\ref{appendix:cogen}.
It is used to construct calls to the unfolder and memoisation 
predicates.
For example, calling {\tt\small add\_extra\_argument("\_u",p(a),C,Code)}
  gives {\tt\small Code = p\_u(a,C)}.
 
Using this more refined treatment, the $\mathit{cogen}$ will produce the 
following unfolder predicate:

{\small
\begin{verbatim}
  map_u(B,[],[],true).
  map_u(C,[D|E],[F|G],(H,I)) :- 
    J =..[C,D,F], add_extra_argument("_u",J,H,K), call(K),
    map_m(C,E,G,I).
\end{verbatim}}

The specialised code obtained for the call {\tt\small map(inc,I,O)} is then:

{\small
\begin{verbatim}
  map__0([],[]).
  map__0([B|C],[D|E]) :-  D is B + 1, map__0(C,E).
\end{verbatim}}

All the overhead of map has been removed and we have even achieved 
unfolding of {\tt\small\small inc}.

In the case we know the length of the list, we can even go 
further and remove the list processing overhead.
In fact, we can now produce the BTC $({\cal A},\Delta')$ with
$\Delta' = \{map(\mathit{static},\mathit{list}(\mathit{dynamic}),\mathit{dynamic}),$ 
 $inc(\mathit{dynamic},\mathit{dynamic})\}$.
If we then specialise
 $map(inc,[X,Y,Z],O)$ we obtain the following:
\begin{verbatim}
  map__0(B,C,D,[E,F,G]) :- E is B + 1, F is C + 1, G is D + 1.
\end{verbatim}

\ignore{**
Let us now return to the case of a call to a primitive (or an externally
defined predicate). For this the $\mathit{cogen}$ does not know the code of the
primitive and the above method will not work. We could go on and introduce
a new annotation for calls to primitives, but this will just bring us in
trouble if a call predicate could call both primitive and predicates. The
way to solve this dilemma is to only allow calls to predicates. This is not
a severe restriction, since programs that can call primitives can be
translated into an equivalent program that does not, as the following
example shows:
\begin{verbatim}
  p(C) :- call(C).
  q :- p(fail).
\end{verbatim}
is translated in to the equivalent program:
\begin{verbatim}
  p(C) :- call(C).
  q :- p(f).
  f :- fail.
\end{verbatim}
**}

\ignore{
\subsection{Multi-Level $\mathit{cogen}$ and Runtime Code Generation}

Instead of pretty-printing assert immediately.

\begin{verbatim}
    rcg_call(Call) :-
       add_extra_argument("_m",Call,SpecCall,MemoCall),
       MemoCall,
       SpecCall.
\end{verbatim}
}

\section{Experimental Results}
\label{section:results}
\label{section:benchmarks}

In this section we present a series
 of detailed experiments with our {\sc logen} system
 as well as with some other specialisation systems.

A first experimental evaluation
 of the $\mathit{cogen}$ approach
 for Prolog was performed
 in \cite{JorgensenLeuschel:Cogen}.
However, due to the limitations of the initial $\mathit{cogen}$
 only very few realistic examples could be analysed.
Indeed, most interesting partial deduction examples
 require the treatment of partially instantiated data, and the initial
 $\mathit{cogen}$ was thus not very useful in practice.
The improved $\mathit{cogen}$ of this paper can now deal with 
 such examples and we were able to
 run our system on a large selection of benchmarks from
 \cite{Leuschel96:ecce-dppd}.
We only excluded those benchmarks
 in \cite{Leuschel96:ecce-dppd} which are specifically tailored
 towards testing tupling or deforestation capabilities
 (such as applast, doubleapp, flip, maxlength, remove,
  rotate-prune, upto-sum, ...),
 as neither {\sc logen} nor {\sc lix} (nor {\sc mixtus}) will
 be able to achieve any interesting specialisation on them. 

To test the ability to specialise non-declarative built-ins 
 we also devised one new non-declarative benchmark:
 specialising the non-ground unification algorithm
  with occurs-check from page 152 of
  \cite{SterlingShapiro:ArtOfProlog} for the query
 {\tt\small unify(f(g(a),a,g(a)),S)}.
More detailed descriptions about all the benchmarks can be found
 in \cite{Leuschel96:ecce-dppd}.

Our new
 {\sc logen} system
 runs under Sicstus Prolog
 and is publicly available at {\tt\small http://www.ecs.soton.ac.uk/\~{}mal}
 (along with the {\sc lix} system).
We compare the results of {\sc logen} with the latest versions
 of {\sc mixtus}
 \cite{Sahlin93:ngc} (version 0.3.6)
 and {\sc ecce}
 \cite{LeuschelMartensDeSchreye:Toplas,CPD:megapaper}.
(Comparisons of the initial $\mathit{cogen}$
 with other systems such as {\sc logimix},
 {\sc paddy}, and {\sc sp} can be found
 in \cite{JorgensenLeuschel:Cogen}).
For evaluation purposes,
 we will also compare with our traditional offline specialiser
 {\sc lix},
 which performs exactly the same specialisation
 as {\sc logen} (and works on exactly the same annotations).
As we have the {\sc logen} at our disposal,
 we have not tried to make {\sc lix}
 self-applicable, although we conjecture that,
 using our extensions developed in Section~\ref{section:cogen-impure},
 it should be feasible to do so (especially since {\sc lix} was
 derived from {\sc logen}).
 
All the benchmarks were run under
 {\tt\small SICStus Prolog 3.7.1} on a Sun Ultra E450 server
  with 256Mb RAM operating under
 {\tt\small SunOS 5.6}.
 
\begin{table}[htb]
\begin{center}\begin{small}
\begin{tabular}{l|r|rr|rr|r} 
Benchmark  & \multicolumn{1}{c|}{\sc mixtus}   
  &  \multicolumn{2}{c|}{\sc ecce} & 
    \multicolumn{2}{c|}{\sc logen} &
 {\sc lix} \\
~  & \multicolumn{1}{c|}{\footnotesize with}    & 
  \multicolumn{1}{c}{\footnotesize with} &
  \multicolumn{1}{c|}{\footnotesize w/o}  & 
  \multicolumn{1}{c}{\footnotesize cogen} &
  \multicolumn{1}{c|}{\footnotesize genex} & ~ \\
\hline
advisor & 70 ms & 50 ms & 20 ms & 2.6 ms & 0.8 ms & 1.7 ms\\
contains.kmp & 210 ms & 550 ms & 400 ms & 1.9 ms & 3.6 ms & 5.6 ms\\
ex{\tt\_}depth & 200 ms & 230 ms & 190 ms & 1.5 ms & 7.2 ms & 7.6 ms\\

grammar & 220 ms & 200 ms & 140 ms & 6.3 ms & 1.0 ms & 1.3 ms\\

groundunify.simple & 50 ms & 50 ms & 20 ms & 6.5 ms & 7.7 ms & 8.9 ms\\
groundunify.complex  & 990 ms & 4080 ms & 3120 ms & "  & 8.0 ms & 9.3 ms\\

imperative-solve ~ & 450 ms & 5050 ms & 4240 ms & 7.3 ms  & 4.3 ms & 9.2 ms\\

map.rev & 70 ms & 60 ms & 30 ms & 2.7 ms & 1.1 ms  & 1.1 ms\\
map.reduce & 30 ms & 60 ms & 30 ms & " & 1.4 ms & 1.4 ms \\

match.kmp & 50 ms & 90 ms & 40 ms & 1.0 ms & 2.5 ms & 2.8 ms\\

model{\tt\_}elim & 460 ms & 240 ms & 170 ms & 3.0 ms & 3.1 ms & 3.3 ms\\

regexp.r1 & 60 ms & 110 ms & 80 ms & 1.3 ms & 1.4 ms & 2.0 ms\\
regexp.r2 & 240 ms & 120 ms & 80 ms & "  & 2.5 ms & 4.1 ms\\
regexp.r3 & 370 ms & 160 ms & 120 ms & "  & 9.9 ms & 17.2 ms\\

ssuply & 80 ms & 120 ms & 60 ms & 5.5 ms & 0.7 ms & 2.4 ms\\
transpose & 290 ms & 190 ms & 150 ms & 1.0 ms & 1.9 ms & 2.3 ms\\
ctl  & 40 ms & 160 ms & 230 ms & 4.4 ms & 1.3 ms & 1.7 ms\\

ng{\tt\_}unify & 2510 ms & {\footnotesize na} & 
\multicolumn{1}{c|}{\footnotesize na} & 5.3 ms & 3.5 ms & 5.7 ms\\

\hline

Total {\tiny (except ng\_unify)} &
  3860 ms &  11520 ms &  9120 ms &  45 ms & 58 ms & 82 ms\\
~~~~~~~~{\scriptsize normalised:} &  66  &  197  &  156  &  0.77  & 1  &
1.40
\\

\end{tabular}
\end{small}
\caption{Specialisation Times\label{table:specialisation}}
\end{center}
\end{table}

\begin{table}[tb]
\begin{center}\begin{small}
\begin{tabular}{l|rrrr} 
Benchmark  & ~~~Original & ~~~~{\sc mixtus}   &  ~~~~~~~{\sc ecce} & ~~{\sc logen}
 / {\sc lix} \\
\hline
advisor  
~ & 1 & 3.94 & 3.29 & 3.94\\
contains.kmp 
~ & 1 & 5.17 & 6.2 & 4.89\\
ex{\tt\_}depth 
~ & 1 & 2.16 & 2.72 & 2.77\\
grammar 
~ & 1 & 14.40 & 9.60 & 15.16\\
groundunify.simple 
~ & 1 & 14.00 & 14.00 & 1.56\\
groundunify.complex~~~~~ 
~ & 1 & 14.33 & 14.33 & 14.33\\
imperative-solve~ 
~ & 1 & 1.35 & 2.56 & 1.35\\
map.rev  
~ & 1 & 2.30 & 1.53 & 1.92\\
map.reduce 
~ & 1 & 3.00 & 3.60 & 3.18\\
match.kmp 
~ & 1 & 1.46 & 1.93 & 1.15\\
model{\tt\_}elim 
~ & 1 & 3.56 & 3.78 & 2.69\\
regexp.r1 
~ & 1 & 6.23 & 4.26 & 6.35\\
regexp.r2 
~ & 1 & 2.50 & 2.57 & 3.00\\
regexp.r3 
~ & 1 & 3.36 & 3.14 & 1.15\\
ssuply 
~ & 1 & 51.00 & 51.00 &  51.00\\
transpose 
~ & 1 & 22.71 & 22.71 & 22.71\\
ctl 
~ & 1 & 5.85 & 5.64 & 5.85\\
ng{\tt\_}unify 
~ & 1 & 4.44 & - & 3.72\\
\hline
Average Speedup & 1 & 9.25 &    8.99    & 8.41 \\
Total Speedup & 1 & 3.63 &      3.89    & 2.83 \\
\end{tabular}
\end{small}
\caption{Speedups of the specialised programs\label{table:speedups}}
\end{center}
\end{table}

\subsubsection*{Specialisation Times}
A summary of all the transformation times can be found in
 Table~\ref{table:specialisation}.
The times for {\sc mixtus} contains the time to write the specialised 
program to file (as we are not the implementors of {\sc mixtus} we 
were unable to factor this part out),
 as does the column marked ``with'' for {\sc ecce}.
The column marked ``w/o'' is the pure transformation time of {\sc 
ecce} without measuring the time needed for writing to file.
The times for {\sc logen} exclude writing to file.
Note that {\sc ecce} can only handle declarative programs, and could 
therefore not be applied on the $\mathit{ng}${\tt\_}$\mathit{unify}$ benchmark.
For {\sc logen},
 the column marked by $\mathit{cogen}$ contains
 the runtimes of the $\mathit{cogen}$
 to produce the generating extension,
 whereas the column marked by $\mathit{genex}$ contains the times needed
 by the generating extensions to produce the specialised programs.
To be fair, it has to be emphasised that the
 binding-type analysis for {\sc logen} and {\sc lix}
 was carried out {\em by hand}.
In a fully automatic system thus, the column with the $\mathit{cogen}$
 runtimes will have to be increased by the time needed for the 
 binding-type analysis. The same is true for the {\sc lix} column.
In general, the binding-type analysis will be
 the most expensive operation in one-shot applications,
 and we will address this issue in more detail in the next section.
However, the binding-type analysis and the
 $\mathit{cogen}$ have to be run only {\em once\/} for every 
 program and division.
For example, the generating extension produced for
 $\mathit{regexp}.r1$ was re-used without modification for
 $\mathit{regexp}.r2$ and $\mathit{regexp}.r3$
 while the one produced for $\mathit{map}.\mathit{rev}$
 was re-used for $\mathit{map}.\mathit{reduce}$.
Another example is the $\mathit{ctl}$ interpreter for computation
 tree logic which is specialised over and over again for
 different systems and different CTL temporal logic formulas, e.g.,
 in \cite{LeuschelLehmann:Coverability}.
Hence, in a context where the same program is specialised
 over and over again for different static values,
 the time devoted to the {\em BTA\/} will usually become negligible.

In summary, the results in this section are valid in a setting
 where a knowledgeable user can produce a good
 and safe {\em BTC\/} by hand
 (we have developed a Tcl/Tk based graphical front end that helps
 the user by providing visual feedback about the annotations)
 and the same program is re-specialised multiple times.
 


As can be seen in Table~\ref{table:specialisation},
 \system{logen} and {\sc lix} are the fastest specialisation 
 systems overall, running up to 
 almost 3 orders of magnitude faster than the existing online systems.
{\sc lix} runs roughly 40 \% slower
 than the generating extensions of {\sc logen}.
Note that for 3 benchmarks ($\mathit{contains}.\mathit{kmp}$,
 $\mathit{regexp}.r2/3$) the cost of running the $\mathit{cogen}$
 is already re-covered after a single specialisation.
All in all, specialisation times of both {\sc logen} and {\sc lix}
 are very satisfactory and seem to be more predictable than that of
 online systems.

\subsubsection*{Quality of the Specialised Code}
Table~\ref{table:speedups} contains the
 speedups obtained by the various systems.
The table also contains the overall average speedup and total speedup.
The latter is a fairer measure
than average speedup and is obtained
by the formula
 $\frac{n}{\sum_{i=1}^{n}\frac{\mathit{spec}_i}{\mathit{orig}_i}}$
 where $n$ is the number of benchmarks
 and $\mathit{spec}_i$ and $orig_i$ are the absolute execution
 times of the specialised and original programs respectively.

As can be seen in Table~\ref{table:speedups}, the 
specialisation performed by the {\sc logen} system is not very far 
off the one obtained by {\sc mixtus} and {\sc ecce}; sometimes {\sc 
logen} even surpasses both of them (for $\mathit{ex}${\tt\_}$\mathit{depth}$,
 $\mathit{grammar}$, $\mathit{regexp}.r1$ and $\mathit{regexp}.r2$).
Being a pure offline system, {\sc logen} cannot pass the KMP-test,
 which can be seen in the timings for $\mathit{match}.\ mathit {kmp}$ in
  Table~\ref{table:speedups}.
(To be able to pass the KMP-test, more sophisticated local control
 would be required, see \cite{MartinLeuschel:PSI99}
  and the discussion below.)

Again, to be fair, both {\sc ecce} and {\sc mixtus} are fully automatic 
systems guaranteeing termination, while for {\sc logen}
 sufficient specialisation
 and termination had to be manually
 ensured by the user via the {\em BTC\/}.
We return to this issue below.
Nonetheless, the {\sc logen} system is surprisingly fast and produces 
surprisingly good specialised programs.

Finally, the figures of {\sc logen} in Tables~\ref{table:specialisation}
and \ref{table:speedups}
 shine when compared to the self-applicable \system{sage}\ system,
 where compiler generation usually takes more
 than $10$ hours (with garbage collection) \cite{Gurr:PHD}
 and where the resulting generating extension are still pretty slow \cite{Gurr:PHD}
 (taking more than $100 000 ms$ to produce the
 specialised program;
  unfortunately self-applying \system{sage}\ is not possible
 for normal users and we cannot make exact comparisons with 
 {\sc logen}).


\section{Automating Binding-time Analysis}
\label{section:bta}


Automating the process of binding-time analysis has received
a lot of attention in the context of functional and imperative languages
\cite{Bondorfetal:jfl93,Consel:1993:PolyvariantBinding-Time}.
In the context of logic programs, 
a major step in achieving automatic binding-time analysis has recently been
the use of termination analysis
\cite{mauriceetal:esop98,Vanhoof01:proc}.
In what follows, we highlight the main aspects of \cite{Vanhoof01:proc}
and report on some experiments.

\subsection{Automatic Binding-time Analysis}

When annotating a program, one generally wants to mark
as many atoms {\em reducible} as possible, while guaranteeing 
termination of the unfolding.  
In order to study the termination characteristics of an unfolding rule
$U_{\mathcal A}$ associated to an annotation ${\mathcal A}$, 
we adopt a slightly different notion of annotation from \cite{Vanhoof01:proc}.
The basic idea is to represent the annotation ${\mathcal A}$ on a clause
by a new clause (which we will call a {\em t-annotation}) 
in which the non-reducible atoms are replaced by $\mathit{true}$. This will allow 
to mimic unfolding using $U_{\mathcal A}$ by normal evaluation of
the corresponding t-annotation.
\begin{definition}\label{def:ann}
Given a clause $H\leftarrow B_1,\ldots,B_n$,  a {\em t-annotated version of}
the clause is a clause $H\leftarrow B_1',\ldots,B_n'$, where for each 
$i$ such that
$1\le i\le n$, it holds that either $B_i' = B_i$ or $B_i' = \mathit{true}$. A t-annotated
version of a program $P = \bigcup_i C_i$ is a program $P' = \bigcup_i C_i'$
such that for every such clause $C_i$, it holds that  $C_i'$ is a t-annotated 
version of $C_i$.
\end{definition}
Note that, according to Definition~\ref{def:ann}, every clause is a t-annotated
version of itself. 
Given an annotation ${\mathcal A}$ for a program $P$, we will denote with
$\overline{P}_{\mathcal A}$ the t-annotated version of $P$ obtained by
replacing the atoms that are marked {\em non-reducible} by ${\mathcal A}$ with
$\mathit{true}$. Note that there is a one-to-one correspondence between ${\mathcal A}$,
$P_{\mathcal A}$ and $\overline{P}_{\mathcal A}$
 and in what follows we will freely switch between them,  referring simply 
to an ``annotated'' program.
The introduction of a t-annotation allows to
reason about the termination behaviour of an unfolding rule
 $U_{\mathcal A}$ when unfolding $P\cup\{G\}$ by studying the 
termination behaviour of $\overline{P}_{\mathcal A}$ with respect to $G$.
Indeed, if $\overline{P}_{\mathcal A}$ terminates for a goal $G$, 
then the (possibly incomplete) SLD-tree for $P\cup\{G\}$ built
by $U_{\mathcal A}$ is finite and vice versa. 

The above observation is the core of the algorithm developed by
\cite{Vanhoof01:proc}, which computes a terminating t-annotation of a 
program $P$ for a goal $G$.
The basic intuition behind the algorithm, which is depicted in
Fig.~\ref{fig:alg}, is as follows:
suppose we have to annotate a program $P$ with respect to an
initial goal $G$. If we can prove that $G$ terminates with respect to $P$,
the t-annotated version of $P$ returned by the algorithm is simply $P$ 
itself (corresponding with a $P_{\mathcal A}$ in which every atom is 
annotated reducible). Hence, $U_{\mathcal A}$ constructs a
complete SLD-tree for $P\cup\{G\}$ and specialisation of $G$ boils down to
plain evaluation.
If, on the other hand, termination of $G$ with respect to the t-annotation
under construction can not be proven by the analysis due to the presence of 
a possible loop, the algorithm tries to remove the loop by replacing an atom
by $\mathit{true}$. This process is repeated until the constructed t-annotation, and
hence the annotated program, is proven to be loop free.  

\newcommand{\acalls}{\ensuremath{calls^\alpha}}
\newcommand{\lla}{\ensuremath{\mbox{{\it LLA}}}}

To characterise the possible loops in a program (or a t-annotation) $P$, 
the analysis first identifies which of the atoms 
are {\em loop-safe}. Intuitively, an atom $B_i$ in a clause 
$H\leftarrow B_1,\ldots,B_n\in P$ is said to be {\em loop safe} if 
the analysis can prove that a finite SLD-tree is built for 
any atom from the program's callset
(the set of calls that can possibly arise during evaluation of $P\cup\{G\}$) 
that unifies with $H$ if the tree is constructed by unfolding only the 
$i$ leftmost body atoms of the clause under
consideration.
Computing whether an atom is loop safe is achieved by known techniques of
termination analysis.
In our work, we followed the approach of \cite{codish99:jlpb}. 
A norm $\|.\|$ is chosen -- mapping a term to a natural
number -- and the program's callset is approximated by
a finite abstract callset, denoted by $\acalls_P(G)$. 
Every call in $\acalls_P(G)$ is of the form $p(b_1,\ldots,b_n)$ with $b_i$ a
boolean stating whether or not the size of that argument (according to the
chosen norm)
can change  upon further instantiation. 
More formally, we can define the generalisation of a call $p(t_1,\ldots,t_n)$ as
$p(\alpha_{\|.\|}(t_1),\ldots,\alpha_{\|.\|}(t_n))$, where
$\alpha_{\|.\|}$ is defined as follows, mapping terms onto the boolean domain
$\{\mathit{false},\mathit{true}\}$ with $\mathit{false} > \mathit{true}$:
\[\alpha_{\|.\|}(t) = \left\{\begin{array}{ll}
\mathit{true} & \mbox{if }\|t\theta\| = \|t\|\mbox{ for any }\theta\\
\mathit{false} & \mbox{otherwise}
\end{array}\right.\]
The abstract callset is kept monovariant -- containing a single call per
predicate --  by taking the predicate-wise least upper bound of the calls in the
set.
The system then concludes loop-safeness of an atom $B_i$ in a clause
$H\leftarrow B_1,\ldots,B_n$ if it can show that there is a guaranteed
decrease in size between $H$ and any recursive call that may occur during
unfolding of $B_1,\ldots,B_{i}$ given the calls in $\acalls_P(G)$ and the
size relations between the sizes of the arguments in $B_1,\ldots,B_{i-1}$.

Given the atoms that are guaranteed to be loop safe, the algorithm identifies
in each of the clauses the leftmost atom -- if it exists -- which is not 
proven to be loop safe, and removes one of these.
\begin{figure}[hbtp]
\centering
\begin{tabular}{l}
Given a program $P$ and initial goal $G$.\\
Let $P_0 = P$, $S_0=\acalls_{P}(G)$, $k=0$.\\
{\bf repeat}\\
\qin if there exist a clause $i$ in $P_{k}$ such that the $j$'th body
atom\\
\qin cannot be proven to be loop-safe given $S_k$\\
\qin then\\
\qin\qin let $P_{k+1}$ be the program obtained by replacing the $j$'th\\ 
\qin\qin body atom in the $i$'th clause in $P_k$ by $\mathit{true}$ and \\
\qin\qin let $S_{k+1} = S_{k}\sqcup\acalls_{P_{k+1}}(G)$\\
\qin else\\
\qin\qin  $P_{k+1} = P_{k}$\\
\qin $k = k+1$\\
{\bf until} $P_k = P_{k-1}$\\
$P'=P_k$, $S'=S_k$\\
\end{tabular}
\caption{The binding-time analysis algorithm.}
\label{fig:alg}
\end{figure}
Note that the algorithm is non deterministic, as several such clauses may
exist. Also note the construction of the set $S'$: starting from the program's
initial abstract callset $S_0$, in each round the predicate-wise 
least upper bound is computed with the current t-annotation's abstract callset.
Doing so
guarantees that the calls that are unfolded are correctly
represented by an abstract call in $S'$, but it also ensures that $S'$
contains abstractions of the (concrete instances
of the) calls that were replaced by $\mathit{true}$ during the process. In other
words, the set $S'$ contains an abstraction of every call that is encountered
(unfolded or residualised) during specialisation of $P$ with respect to the
initial goal $G$. 
Termination of the algorithm is straightforward, since in every
iteration an atom in a clause is replaced by $\mathit{true}$, and the program
only has a finite number of atoms. 

\begin{example}\label{ex:vanilla}
Consider the meta interpreter depicted in Fig.~\ref{fig:vanilla}. The
interpreter has the {\tt member/2} and {\tt append/3} predicates as
object program.

\begin{figure}[hbtp]\small
\centering
\begin{tabular}{@{}l}
\begin{SProg}
1:solve([]).\\
2:solve([A|Gs]):- solve\_atom(A), \underline{solve}(Gs).\\
\\
3:solve\_atom(A):-\underline{clause}(A,Body), \underline{solve}(Body).\\
\\
4:clause(member(X,Xs), [append(\_,[X|\_],\_Xs)]).\\
5:clause(append([],L,L), []).\\
6:clause(append([X|Xs],Y,[Z|Zs]),[append(Xs,Y,Zs)]).\\
\end{SProg}
\end{tabular}
\caption{Vanilla meta interpreter}
\label{fig:vanilla}
\end{figure}
\end{example}

\noindent
The binding-time analysis inherits from its underlying termination analysis
\cite{codish99:jlpb} 
the need for a norm to be selected by the user. 
An often used norm on values of the type $list(T)$ is the so-called {\em
listlength} norm, counting the number of elements in a list. It is defined as
follows:
\[\begin{array}{lll}
\|\,[\,]\,\| & = & 0\\
\|\,[\_\,|\,Xs]\,\| & = & 1 + \|Xs\|\\
\end{array}\]
Running the binding-time analysis of \cite{Vanhoof01:proc} on the 
program depicted in Example~\ref{ex:vanilla} with respect to the 
listlength norm and the initial goal {\tt solve([mem(X,Xs)])} results in an
annotated program in which the call to {\tt solve\_atom/1} is annotated {\em
non-reducible} and every other call as {\em reducible}. The resulting abstract
callset is
\[\{ solve(\mathit{true}), solve\_atom(\mathit{false}), clause(\mathit{false},\mathit{false})\}\]
denoting that every call to {\tt solve/1} has an argument that is at least
bound to a list skeleton, whereas 
the arguments in calls to {\tt solve\_atom/1} and {\tt clause/2} may be of
any instantiation.

Note that there is a close correspondence between the abstract callset and a
(monovariant) division. If we define the concretisation function 
$\gamma_{\|.\|}$ mapping a boolean to a type as 
$\gamma_{\|.\|}(b) = \tau$
where $\tau$ is the most general type such that for all terms 
$t:\tau$ holds that $\alpha_{\|.\|}(t) \le b$, 
then we can define the division corresponding
to an abstract callset $S$ as
\[\Delta = \{p(\gamma_{\|.\|}(b_1),...\gamma_{\|.\|}(b_n))\:|\:
p(b_1,\ldots,b_n)\in S\}.\]
If $U_{\mathcal A}$ and $\Delta$ represent, respectively,
the unfolding rule and the division corresponding with the t-annotation and
abstract callset computed by the binding-time algorithm, then 
$(U_{\mathcal A},\Delta)$ is a globally safe binding-time
classification for the program under consideration. 
\begin{example}
The division corresponding with the callset above is
\[\Delta = \{solve(\mbox{{\tt list(dynamic)}}), solve\_atom(\mbox{{\tt dynamic}}),
clause(\mbox{{\tt dynamic}})\}\]
where the parametric type {\tt list(.)} is defined as before:
\begin{verbatim}
:- type list(T) ---> [ ] ; [T | list(T)].
\end{verbatim}
\end{example}


\subsection{Additional Experiments}


Table~\ref{table:bta:benchmarks} summarises a number of experiments that 
were run with the
binding-time analysis of \cite{Vanhoof01:proc}.  
We could not use all of the benchmarks from Section~\ref{section:benchmarks},
 because the current {\em BTA\/} is not yet capable of treating
 some of the built-ins required and,
 while it can handle partially static data, it can only
 handle one kind of partially
  static data (depending on the single norm with respect to which the
  program is analysed).

The timings in Table~\ref{table:bta:benchmarks} are in
milliseconds and were measured on the same machine and Prolog system used in
Section~\ref{section:results}. 
The second column ({\em Round1}) presents the timings for
termination analysis of the original program (in which all calls are annotated
reducible). In case the outcome of the analysis is possible non-termination,
the third column presents the timings for termination analysis of the program
from which a call was removed. None of the benchmarks required more than two
rounds of the algorithm to derive a terminating t-annotation. The fourth
column then contains the total time needed to produce the generating extension
using {\sc logen} and to run it on the partial deduction query. The final
column contains the specialisation time of {\sc mixtus} (from
Section~\ref{section:results}) as a reference point.

\begin{table}[h]
\[\begin{array}{l|rr|r|r||r} 
\multicolumn{1}{l|}{\mbox{Benchmark}} &
\multicolumn{1}{c}{\mbox{Round 1}} &
\multicolumn{1}{c|}{\mbox{Round 2}} &
\multicolumn{1}{c|}{\mbox{{\sc logen}}} &
\multicolumn{1}{c||}{\mbox{Total}} & \mbox{{\sc mixtus}}
\\ \hline
\mbox{{ ex{\tt\_}depth}} &
240.0 \mbox{ ms} &
230.0 \mbox{ ms} & 4.4 \mbox{ ms} &
474 \mbox{ ms} &
200 \mbox{ ms}
\\
\mbox{{ match.kmp}} &
470.0 \mbox{ ms} &
180.0 \mbox{ ms} & 2.4 \mbox{ ms} &
652 \mbox{ ms} &
50 \mbox{ ms}\\
\mbox{{ map.rev/reduce}} &
200.0 \mbox{ ms} &
\mbox{--} & 4.3 \mbox{ ms} &
204 \mbox{ ms} &
100 \mbox{ ms}\\
\mbox{{ regexp.r1-3}} &
740.0 \mbox{ ms} & 
280.0 \mbox{ ms} & 15.1 \mbox{ ms} &
1035 \mbox{ ms} &
670 \mbox{ ms}\\
\mbox{{ transpose}} &
210.0 \mbox{ ms} &
150.0 \mbox{ ms} & 7.0 \mbox{ ms} &
367 \mbox{ ms} &
290 \mbox{ ms}\\
\hline
\mbox{{ Total}} &
\multicolumn{2}{c|}{2850 \mbox{ ms}} & 34.5 \mbox{ ms} &
2885 \mbox{ ms} &
1330 \mbox{ ms}\\
\end{array}\]
\caption{Timings for the binding-time analysis and
 full specialisation.}
\label{table:bta:benchmarks}
\end{table}

Note that we slightly modified the {\em transpose} benchmark in the sense that
the first argument is fully static. In fact, in the original {\em transpose}
benchmark the first argument is a list skeleton whose first element in turn is
a list skeleton but whose other elements are dynamic. This binding-type cannot
be represented precisely by a semi-linear norm (which is required by the
termination analysis of \cite{codish99:jlpb} underlying the 
binding-time analysis).

Analysing Table~\ref{table:bta:benchmarks} 
we can see that the binding-time analysis is indeed
the most expensive operation in a one-shot situation. However, the timings are
not too bad compared to {\sc mixtus} and the cost of the binding-time analysis
will already be recovered after a few specialisations (e.g., after 3
iterations for {\em ex\_depth} and after 2 iterations for {\em regexp.r3}).
%
Table~\ref{table:speedups2} 
contains a summary of the speedups obtained by the {\sc
logen} (or {\sc lix}) system when using the annotations obtained by the above
binding-time analysis. For comparison's sake we have also added the
corresponding speedups using the methods of Section~\ref{section:results}.
Observe that,
as was probably to be expected, the automatically generated annotations lead to less
speedups than using hand-crafted annotations. 
Indeed, the hand-crafted annotations for {\em ex\_depth}
 uses the {\tt hide\_nf} annotation to prevent duplication
of expensive calls as described in Section~\ref{section:expensive_pred},
 the hand-crafted annotations for {\em map.rev} and {\em map.reduce}
 uses the special annotations for the {\tt call} primitive described in
 Section~\ref{section:call-pred}, while for {\em transpose}
 the termination analysis of the automatic {\em BTA} classified one
 call as non-terminating which is in fact terminating.
Nonetheless, the figures are
still pretty good, for the 3 {\em regexp} benchmarks we obtain exactly the
same result as the hand-crafted annotation and for the {\em match.kmp} the
automatic annotation actually outperforms the hand-crafted one.

\begin{table}[tb]
\begin{center}\begin{small}
\begin{tabular}{l|rrrr|r}
Benchmark  & Original & {\sc mixtus}   &  {\sc ecce} & {\sc logen}
 hand  & ~{\sc logen} automatic\\
\hline
ex{\tt\_}depth 
~ & 1 & 2.16 & 2.72 & 2.77 & 2.23\\ 

map.rev  
~ & 1 & 2.30 & 1.53 & 1.92 & 1.53\\

map.reduce 
~ & 1 & 3.00 & 3.60 & 3.18 & 1.29\\

match.kmp 
~ & 1 & 1.46 & 1.93 & 1.15 & 1.34\\

regexp.r1 
~ & 1 & 6.23 & 4.26 & 6.35 & 6.35\\

regexp.r2 
~ & 1 & 2.50 & 2.57 & 3.00 & 3.00\\

regexp.r3 
~ & 1 & 3.36 & 3.14 & 1.15 & 1.15\\

transpose 
~ & 1 & 22.71 & 22.71 & 22.71 & 5.89\\

\hline
Average    Speedup~~ & 1 & 5.47    & 5.31    & 5.28    & 2.85\\
Total    Speedup & 1 & 2.84    & 2.85    & 2.31    & 1.93\\

\end{tabular}
\end{small}
\caption{Speedups of the specialised programs\label{table:speedups2}}
\end{center}
\end{table}

The conducted experiments show that the approach is feasible and can be
automated.
However, some issues regarding the current binding-time analysis remain. 
The analysis basically deals with boolean binding-times: either a value is
instantiated enough with respect to a norm, or it is not. 
%
Recent research \cite{TypedNorms,Vanhoof:Lopstr01}
shows that termination proofs can be constructed by measuring the size
of a term by means of a number of simple norms rather than using a single
sophisticated norm. These simple norms basically count the number 
of subterms of the term that are of a particular type.
In the presence of type information these norms can be constructed
automatically. 
When combined with information that denotes whether
further instantiating a term can introduce more subterms of the particular
type 
they provide a more fine-grained characterisation of a term's size 
and instantiation.
We conjecture such a more detailed characterisation 
to be a powerful and promising mechanism to derive an automatic
binding-time analysis capable of constructing more precise binding-types. 
%
Also note that the current analysis only produces monovariant divisions.
Polyvariance of the analysis can in principle 
be obtained by allowing several calls to the
same predicate in the abstract callset, 
creating a new variant of the predicate definition for each abstract call and
checking termination of each such predicate separately.
%

In summary, at least for the experiments in Tables~\ref{table:bta:benchmarks}
and \ref{table:speedups2}, we can conclude that online systems are to be
preferred -- both in terms of speed and quality of the specialised code -- in 
one-shot situations where no expert user is available to perform the 
annotation. Nonetheless, the quality of the fully automatic {\sc logen}
is satisfactory and the cost of the binding-time analysis will usually be
recovered already after a few specialisations.
This means that the fully automatic {\sc logen} might be useful in situations
were the same program is specialised multiple times and the specialisation
times itself are of utmost importance. Further work is needed to extend and refine the binding-time analysis and to establish its scaling properties 
for larger programs.


\section{Discussion and Future Work}
\label{sec-related-work}
\label{section:discussion}

%
%

\subsection{Related Work}
\label{section:related-work}

The first hand-written compiler generator based on partial evaluation
principles was, in all probability, the system {\em RedCompile\/}
 \cite{Beckman:76} for a
dialect of Lisp. Since then successful compiler
generators have been written for many different languages and language
paradigms
\cite{Romanenko:88,Holst:89a,HolstLaunchbury:92,%
BirkedalWelinder:94:handcogen,Andersen94:PhD,GlueckJoergensen:PLILP95,%
Thiemann:ICFP96}.

In the context of definite clause grammars and parsers based on them, the
  idea of hand writing the compiler generator has also been used in
 \cite{Neumann:META90,Neumann:LPAR90}. 
However, it is not based on (off-line) partial deduction.

Also the construction of our program $P_{u}^{\cal A}$ 
(Definition~\ref{def:pul-transf})
 is related to the idea of {\em abstract compilation}
 \cite{Hermenegildo92:jlp,CodishDemoen:jlp95}.
In abstract compilation 
 a program $P$ is first transformed and abstracted.
Evaluation of
 this transformed program corresponds to the actual 
 abstract interpretation analysis of $P$.
In our case concrete execution of $P_{u}^{\cal A}$ performs
 (part of) the partial deduction process.
Another similar idea has also been
 used in \cite{TarauDeBosschereDemoen:LOPSTR93}
 to calculate abstract answers.
Finally, \cite{GallagherLafave:PE96} uses a source-to-source transformation
 similar to ours to compute trace terms for the global control of
 logic and functional program specialisation (however, the 
 specialisation technique itself is still basically online).

The local control component of our generating extensions is still
rather limited: either a call is always reducible or never reducible.
To remedy this problem, and to allow any kind of partially 
instantiated data, an extension of our cogen approach
 has been developed in \cite{MartinLeuschel:PSI99}.
This approach uses a sounding analysis (at specialisation time)
 to measure the minimum depth of 
 partially instantiated terms. The result of this analysis is
 then used to control the unfolding and ensure termination.
This approach allows more aggressive unfolding 
 than the technique presented in this paper, passing the KMP-test and
 rivalling online systems 
 in terms of flexibility.
Due to the sounding analysis, however, it is not fully offline.
In terms of speed of the specialisation process, it is hence slower 
than our fully offline cogen approach (but still much faster than 
online systems such as {\sc mixtus} or {\sc ecce}).
Also, \cite{MartinLeuschel:PSI99} only addresses the local control 
component and it is still unclear how it can be extended for the 
global control (the prototype in \cite{MartinLeuschel:PSI99}
 uses the online {\sc ecce} system for global control; to this end
 trace terms were built up in the generating extension
 like in \cite{GallagherLafave:PE96}).

Although our approach is closely related to the one for functional
programming languages there are still some important differences.  Since
computation in our cogen is based on unification, a variable is not forced
to have a fixed binding time assigned to it.
In fact the binding-time
analysis is only required to be safe, and this does not enforce this
restriction. Consider, for example, the following program:

{\footnotesize
\begin{verbatim}
  g(X) :- p(X),q(X)
  p(a).   q(a).
\end{verbatim}
}

\noindent If the initial division $\Delta_0$ states that the argument to
\svbt{g} is dynamic, then $\Delta_0$ is safe for the program and the
unfolding rule that unfolds predicates \svbt{p} and \svbt{q}. The residual
program that one gets by running the generating extensions is:

{\footnotesize
\begin{verbatim}
  g__0(a).
\end{verbatim}
}
 
\noindent In contrast to this any cogen for a functional language known to
us will classify the variable \svbt{X} in the following analogous functional
program (here exemplified in Scheme) as dynamic:

{\footnotesize
\begin{verbatim}
  (define (g X) (and (equal? X a) (equal? X a)))
\end{verbatim}
}

\noindent and the residual program would be identical to the original
program.

One could say that our system allows divisions that are not
{\em uniformly\/} congruent in the sense of
Launchbury
\cite{Launchbury:91:book} and essentially, our system
performs specialisation that a partial evaluation system for a functional
language would need some form of {\em driving\/}
\cite{GlueckSorensen:PLILP94} to be able to do.
However, our divisions are still congruent: the
value of a static variable cannot depend on a dynamic value.
In the above example, the value of $X$ within the call $q(X)$,
 if reached,
 is always going to be $a$, no matter what the argument
 to \svbt{g} is.

\subsection{Mixline Specialisation}
\label{subsection:mixline}

Some built-ins can be treated in a more refined fashion than described
 in Section~\ref{section:cogen-impure}.
For instance, for a call {\tt\small var(X)} which is non-reducible we could 
still check whether the call fails or succeeds in the generating extension.
If the call fails, we know that it will definitely fail at runtime as 
well. In that case we don't have to generate code and we thus 
achieve improved specialisation over a purely offline approach.
If the call {\tt\small var(X)} succeeds, however, we have gained nothing
and still have to perform {\tt\small var(X)} at runtime.

Similarly, for a call such as {\tt\small ground(X)}, if it succeeds in the 
generating extension we can simply generate {\tt\small true} in the 
specialised program. In that case we have again improved the 
efficiency of the specialised program.
If, on the other hand,
 {\tt\small ground(X)} fails in the generating extension
 it might still succeed at runtime: 
we have to generate the code {\tt\small ground(X)} and have gained nothing.

The rules below cater for such a mixline
 \cite{Jones:peval} treatment of some built-ins.

$$
\begin{array}{ll}
    \underline{c} \specialise c : c &
      \mbox{if~}c = \mbox{\tt\small var}(t), \mbox{\tt\small copy\_term}(s,t),
      s\backslash\mbox{\tt\small ==}t,\ldots
\\[1ex]
    \underline{c} \specialise ( c \rightarrow C=\mathit{true} ;  C=c) : C &
      \mbox{if~}c = \mbox{\tt\small ground}(t), \mbox{\tt\small nonvar}(t),
      \\
~      & ~~~~~~~~\mbox{\tt\small atom}(t), \mbox{\tt\small integer}(t),
      s\mbox{\tt\small ==}t,\ldots
\end{array}
$$

The code of the $\mathit{cogen}$ in Appendix~\ref{appendix:cogen}
 uses these optimisations if a
 {\tt\small mixcall} annotation is used (these annotations have not been 
 used for the experimental results in Section~\ref{section:benchmarks}).
It also contains a mixline conditional, which reduces the 
conditional to the then branch (respectively else branch) if the test
 definitely succeeds (respectively definitely fails) in the 
 generating extension.

Similarly, one can also produce a new binding-type,
 called {\tt\small mix}, which lies in between {\tt\small static}
 and {\tt\small dynamic} \cite{Jones:peval}.
Basically, {\tt\small mix} behaves like {\tt\small static} for 
 the generalisation $\mathit{gen}_{\Delta}$
 (Definition~\ref{def:gen-delta}) but like {\tt\small dynamic} for filtering
 $\mathit{filter}_{\Delta}$
  (Definition~\ref{def:filter-delta}).
The former means that an argument marked as {\tt\small mix} will not be 
abstracted away by $\mathit{gen}_{\Delta}$, while the latter allows such an 
argument to contain variables.
Again, the code for these improvements can be found in 
Appendix~\ref{appendix:cogen}.

Another worthwhile improvement is to enable {\em mixline\/} unfolding of 
predicates.
In other words, instead of either always or never unfolding a 
predicate, one would like to either unfold the predicate or not based 
upon some (simple) criterion. 
This improvement can be achieved,  
without having to change the $\mathit{cogen}$ itself, by
 modifying the annotation process.
Indeed, instead of marking a call $p(t_{1},\ldots,t_{n})$ either 
 as reducible or non-reducible we simply insert a
 static conditional into the annotated program:
 {\tt\small ($\mathit{Test}$ -> $\underline{p}(t_{1},\ldots,t_{n})$ ; $p(t_{1},\ldots,t_{n})$)}.
Thus, if $\mathit{Test}$ succeeds the generating extension will unfold the 
call, otherwise it will be memoised.
 
We have actually used these improvements to produce a mixline
 annotation of the $\mathit{match}.\mathit{kmp}$ benchmark from 
 Section~\ref{section:benchmarks}.
The results of this experiment (after some very simple 
post-processing) is as follows.
 
\begin{center}
\begin{small}
\[\begin{array}{l|rrrr} 
\mbox{Program}  & \mbox{cogen} & \mbox{genex}   &  
\mbox{spec.\ runtime} & \mbox{speedup} \\
\hline

\mbox{match.kmp} & \mbox{1.2 ms} & \mbox{3.7 ms} & \mbox{2480 ms} & \mbox{1.51}
\times \\
\end{array}\]
\end{small}
\end{center}

Note that {\sc logen} now outperforms {\sc mixtus}, passes the 
KMP-test (actually, even without the
 post-processing; see \cite{Sorensen:PE98}).

\ignore{

[Ideally: use dotted underline below:]
$$
\begin{array}{ll}
    \underline{\underline{var(X)}} \specialise var(X) : var(X)\\
    \underline{\underline{copyterm(X,Y)}} \specialise
        copyterm(X,Y) : copyterm(X,Y)\\
    \underline{\underline{X\backslash==Y}} \specialise X\backslash==Y : X\backslash==Y\\[2ex]
    \underline{\underline{ground(X)}}
        \specialise ( ground(X) \rightarrow C=\mathit{true} ;  C=ground(X)) : C\\
    \underline{\underline{\mathit{nonvar}(X)}}
        \specialise ( \mathit{nonvar}(X) \rightarrow C=\mathit{true} ; 
C=\mathit{nonvar}(X)) : C\\
    \underline{\underline{X==Y}}
        \specialise ( X==Y \rightarrow C=\mathit{true} ;  C=(X==Y)) : C\\
\end{array}
$$
}

\subsection{More Future Work} 

In addition to extending our {\em BTA\/} to generate {\tt\small hide\_nf}
 annotations and to fully integrate the {\em BTA\/} into the
  {\sc logen} system,
 one might also think of further extending its capabilities and domain of 
 application.

First, one could try to extend the
  $\mathit{cogen}$ approach so that it can achieve multi-level
  specialisation \`{a} la \cite{GlueckJoergensen:PLILP95}.
One could also try to use the  $\mathit{cogen}$
 for run time code generation.
A first version of the latter has in fact already been implemented;
 this actually does not require all that many modifications to our
 $\mathit{cogen}$.
The former also seems to be reasonably straightforward to achieve.

Another interesting recent development is 
 fragmental specialisation
 \cite{HelsenThiemann:SAIG00}, where the idea is to
 specialise fragments of the code (such as modules) in the order
in which they arrive.
It should be possible to add such a capability to our $\mathit{cogen}$,
 by using co-routining features (e.g., of SICStus Prolog)
 so as to suspend, for predicates $p$ defined in other fragments,
 calls to the corresponding
 $p_u$ or $p_m$ predicates until the fragment defining
 $p$ is available.

One might also investigate
 whether the $\mathit{cogen}$ approach can be ported to
 other logical programming languages.
It seems
 essential that such languages have some metalevel built-in predicates, like
Prolog's \svbt{findall} and \svbt{call} predicates, for the method to be
 efficient. 
Further work is needed to
 establish whether it is possible
 to adapt the $\mathit{cogen}$ approach
 for G{\"o}del
\cite{HillLloyd:Goedel}
 or Mercury \cite{SomogyiHendersonConway:jlp} so that it still 
 produces efficient generating extensions.

Finally, it also seems natural to investigate to
 what extent more powerful
 control techniques (such as characteristic trees
 \cite{Gallagher91:ngc,LeuschelMartensDeSchreye:Toplas},
  trace terms \cite{GallagherLafave:PE96}
 or the local control of
 \cite{MartinLeuschel:PSI99})
 and specialisation techniques
(like conjunctive partial deduction
 \cite{LeuschelDeSchreyeDeWaal:JICSLP96,%
GlueckJorgensenMartensSorenson:control,CPD:megapaper})
 can be incorporated into the $\mathit{cogen}$, while keeping its 
 advantages in terms of efficiency.

\subsection{Conclusion}

In the present paper we have formalised
 the concept of a {\em binding-type analysis\/}, allowing the 
  treatment of {\em partially static\/} structures,
  in a (pure) logic programming setting and how to
  obtain a generic procedure for offline partial deduction from such an 
  analysis.
We have then
  developed the $\mathit{cogen}$ approach for offline specialisation,
   reaping the benefits of self-application without having to write a 
   self-applicable specialiser.
The resulting system, called {\sc logen}, is surprisingly compact
 and can handle
 partially static data structures,
 declarative and non-declarative built-ins, disjunctions,
 conditionals, and negation.
We have shown that the resulting system achieves fast 
 specialisation
  in situations where the same program
 is re-specialised multiple times.
We have also overcome several limitations of earlier offline systems
 and shown that {\sc logen} can be
applied on a wide range of
  natural logic programs and that the resulting specialisation is also
very good, sometimes even surpassing that of existing online systems.
We have also developed the foundation for
 a fully automatic binding-type
 analysis for the {\sc logen} system, and have evaluated its
 performance on several examples.


\section*{Acknowledgements}

We thank Michael Codish, Bart Demoen, Danny De Schreye,
 Andr\'{e} de Waal, Robert Gl\"{u}ck, Gerda Janssens,
 Neil Jones, Bern Martens,
 Torben Mogensen, Ulrich Neumerkel, Kostis Sagonas, and Peter Thiemann
 for interesting discussions and contributions on this work.
My thanks also go to Laksono Adhianto for developing the
 graphical user interface.
Finally, we are very grateful to anonymous referees for their helpful comments
 and constructive criticism.




\appendix


\section{The Prolog cogen}
\label{appendix:cogen}

This appendix contains the listing of the cogen.
It works on an annotated version of the program to be specialised
 which contains definitions for the following predicates:
\begin{itemize}
\item {\tt\small residual}:
  defines the predicates by which the generating extension is
  to be called, as well as the predicates which are residualised.

\item {\tt\small filter}: the division for the residual predicates

\item {\tt\small ann\_clause}: the annotated clauses where calls in the
 body are annotated by:
\begin{itemize}
 \item {\tt unfold} for reducible user-defined predicates,
 and {\tt memo} for non-reducible user-defined predicates,
 \item {\tt call} for reducible primitives (i.e., built-ins or open predicates; c.f., Section~\ref{section:cogen-impure}),
and  {\tt rescall} for non-reducible user-defined predicates,
 \item {\tt semicall} for non-reducible primitives to be specialised in a mixline fashion (c.f., Section~\ref{subsection:mixline}),
 \item {\tt ucall} for a {\tt call} primitive calling a reducible user-defined predicate and
 {\tt mcall} for a {\tt call} primitive calling a non-reducible user-defined predicate (c.f., Section~\ref{section:call-pred}),
\item {\tt if} and {\tt resif} for reducible and non-reducible conditionals respectively,
 and {\tt semif} for conditionals to be specialised in a mixline fashion
 (c.f., Section~\ref{subsection:mixline}),
\item {\tt not} and {\tt resnot} for reducible and non-reducible negations respectively,
\item {\tt ;} and {\tt resdisj} for reducible and non-reducible disjunctions respectively,
 \item  {\tt hide}, {\tt hide\_nf} to prevent the propagation of bindings and failure.
\end{itemize}
\end{itemize}
An example annotated file can be found in
 Appendix~\ref{appendix:parser}.

{\footnotesize
\begin{verbatim}
/* ----------- */
/*  C O G E N  */
/* ----------- */
:- ensure_consulted('pp').

cogen :-
  findall(C,memo_clause(C),Clauses1),
  findall(C,unfold_clause(C),Clauses2),
  pp(Clauses1),
  pp(Clauses2).

memo_clause(clause(Head,(find_pattern(Call,V) ->
                          true ;
                          (insert_pattern(GCall,Hd),
                           findall(NClause,
                                   (RCall, NClause = clause(Hd,Body)),
                                   NClauses),
                           pp(NClauses),
                           find_pattern(Call,V))) )) :-
  residual(Call), cogen_can_generalise(Call), generalise(Call,GCall),
  add_extra_argument("_u",GCall,Body,RCall),
  add_extra_argument("_m",Call,V,Head).

memo_clause(clause(Head,(find_pattern(Call,V) ->
                          true ;
                          (generalise(Call,GCall),
                           add_extra_argument("_u",GCall,Body,RCall),
                           insert_pattern(GCall,Hd),
                           findall(NClause,
                                   (RCall, NClause = clause(Hd,Body)),
                                   NClauses),
                           pp(NClauses),
                           find_pattern(Call,V))
                        ) )) :-
  residual(Call), not(cogen_can_generalise(Call)),
  add_extra_argument("_m",Call,V,Head).

unfold_clause(clause(ResCall,FlatResBody)) :-
  ann_clause(_,Call,Body),
  add_extra_argument("_u",Call,FlatVars,ResCall),
  body(Body,ResBody,Vars), flatten(ResBody,FlatResBody), flatten(Vars,FlatVars).

body((G,GS),GRes,VRes) :-
  body(G,G1,V),filter_cons(G1,GS1,GRes,true),
  filter_cons(V,VS,VRes,true), body(GS,GS1,VS).

body(unfold(Call),ResCall,V) :- add_extra_argument("_u",Call,V,ResCall).
body(memo(Call),AVCall,VFilteredCall) :-
        add_extra_argument("_m",Call,VFilteredCall,AVCall).

body(true,true,true).
body(call(Call),Call,true).
body(rescall(Call),true,Call).
body(semicall(Call),GenexCall,ResCall) :-
        specialise_imperative(Call,GenexCall,ResCall).

body(if(G1,G2,G3),    /* Static if: */
     ((RG1) -> (RG2,(V=VS2)) ; (RG3,(V=VS3))), V) :-
        body(G1,RG1,_VS1), body(G2,RG2,VS2), body(G3,RG3,VS3).
body(resif(G1,G2,G3), /* Dynamic if: */
     (RG1,RG2,RG3), /* RG1,RG2,RG3 shouldn't fail and be determinate */
     ((VS1) -> (VS2) ; (VS3))) :-
        body(G1,RG1,VS1), body(G2,RG2,VS2), body(G3,RG3,VS3).
body(semif(G1,G2,G3), /* Semi-online if: */
     (RG1,flatten(VS1,FlatVS1),
      ((FlatVS1 == true)
        -> (RG2,SpecCode = VS2)
        ;  ((FlatVS1 == fail)
            -> (RG3,SpecCode = VS3)
            ;  (RG2,RG3, (SpecCode = ((FlatVS1) -> (VS2) ; (VS3))))
           )
     )), SpecCode) :-
 /* RG1,RG2,RG3 shouldn't fail and be determinate */
        body(G1,RG1,VS1), body(G2,RG2,VS2), body(G3,RG3,VS3).

body(resdisj(G1,G2),(RG1,RG2),(VS1 ; VS2)) :- /* residual disjunction */
        body(G1,RG1,VS1), body(G2,RG2,VS2).
body( (G1;G2), ((RG1,V=VS1) ; (RG2,V=VS2)), V) :- /* static disjunction */
        body(G1,RG1,VS1), body(G2,RG2,VS2).
        
body(not(G1),\+(RG1),true) :- body(G1,RG1,_VS1).
body(resnot(G1),RG1,\+(VS1)) :- body(G1,RG1,VS1).

body(hide_nf(G1),GXCode,ResCode) :- 
       (body(G1,RG1,VS1)->
         (flatten(RG1,FlatRG1), flatten(VS1,FlatVS1),
            GXCode = (varlist(G1,VarsG1), 
                      findall((FlatVS1,VarsG1),FlatRG1,ForAll1),
                      make_disjunction(ForAll1,VarsG1,ResCode)));
           (GXCode = true, ResCode=fail)). 
body(hide(G1),GXCode,ResCode) :- 
       (body(G1,RG1,VS1)->
         (flatten(RG1,FlatRG1), flatten(VS1,FlatVS1),
            GXCode = (varlist(G1,VarsG1), 
                      findall((FlatVS1,VarsG1),FlatRG1,ForAll1),
                      ForAll1 = [_|_], /* detect failure */
                      make_disjunction(ForAll1,VarsG1,ResCode)));
           (GXCode = true, ResCode=fail)). 
         
/* some special annotations: */
body(ucall(Call), (add_extra_argument("_u",Call,V,ResCall), call(ResCall)), V).
body(mcall(Call), (add_extra_argument("_m",Call,V,ResCall), call(ResCall)), V).

make_disj([],fail).
make_disj([H],H) :- !.
make_disj([H|T],(H ; DT)) :- make_disj(T,DT).
          
make_disjunction([],_,fail).
make_disjunction([(H,CRG)],RG,FlatCode) :- 
        !,simplify_equality(RG,CRG,EqCode), flatten((EqCode,H),FlatCode).
make_disjunction([(H,CRG)|T],RG,(FlatCode ; DisT)) :-
        simplify_equality(RG,CRG,EqCode), make_disjunction(T,RG,DisT),
        flatten((EqCode,H),FlatCode).

specialise_imperative(Call,Call,Call) :- varlike_imperative(Call),!.
specialise_imperative(Call, (Call -> (Code=true) ; (Code=Call)), Code) :-
         groundlike_imperative(Call),!.
specialise_imperative(X,true,X).

varlike_imperative(var(_X)).
varlike_imperative(copy_term(_X,_Y)).
varlike_imperative((_X\==_Y)).
groundlike_imperative(ground(_X)).
groundlike_imperative(nonvar(_X)).
groundlike_imperative(_X==_Y).
groundlike_imperative(atom(_X)).
groundlike_imperative(integer(_X)).

generalise(Call,GCall) :-
    ((filter(Call,ArgTypes), Call =.. [F|FArgs],
      l_generalise(ArgTypes,FArgs,GArgs))
    -> (GCall =..[F|GArgs])
    ;  (print('*** WARNING: unable to generalise: '), print(Call),nl,
        GCall = Call) ).
 
cogen_can_generalise(Call) :-
    filter(Call,ArgTypes),
    static_types(ArgTypes). /* check whether we can filter at cogen time */

/* types which allow generalisation/filtering at cogen time */
static_types([]).
static_types([static|T]) :- static_types(T).
static_types([dynamic|T]) :- static_types(T).

generalise(static,Argument,Argument).
generalise(dynamic,_Argument,_FreshVariable).
generalise(free,_Argument,_FreshVariable).
generalise(nonvar,Argument,GenArgument) :-
    nonvar(Argument), Argument =.. [F|FArgs],
    make_fresh_variables(FArgs,GArgs), GenArgument =..[F|GArgs].
generalise((Type1 ; _Type2),Argument,GenArgument) :-
    generalise(Type1,Argument,GenArgument).
generalise((_Type1 ; Type2),Argument,GenArgument) :-
    generalise(Type2,Argument,GenArgument).
generalise(type(F),Argument,GenArgument) :-
    typedef(F,TypeExpr), generalise(TypeExpr,Argument,GenArgument).
generalise(struct(F,TArgs),Argument,GenArgument) :-
    nonvar(Argument), Argument =.. [F|FArgs],
    l_generalise(TArgs,FArgs,GArgs), GenArgument =..[F|GArgs].
generalise(mix,Argument,Argument). /* treat as static for generalisation */

l_generalise([],[],[]).
l_generalise([Type1|TT],[A1|AT],[G1|GT]) :-
    generalise(Type1,A1,G1), l_generalise(TT,AT,GT).

make_fresh_variables([],[]).
make_fresh_variables([_|T],[_|FT]) :- make_fresh_variables(T,FT).

typedef(list(T),(struct([],[]) ; struct('.',[T,type(list(T))]))).
typedef(model_elim_literal,(struct(pos,[nonvar]) ; struct(neg,[nonvar]))).

add_extra_argument(T,Call,V,ResCall) :-
  Call =.. [Pred|Args],res_name(T,Pred,ResPred),
  append(Args,[V],NewArgs),ResCall =.. [ResPred|NewArgs].

res_name(T,Pred,ResPred) :-
  name(PE_Sep,T),string_concatenate(Pred,PE_Sep,ResPred).

filter_cons(H,T,HT,FVal) :-
        ((nonvar(H),H = FVal) -> (HT = T) ; (HT = (H,T))).
\end{verbatim}}


\section{The Parser Example} \label{appendix:parser}

The annotated program looks like:

{\footnotesize
\begin{verbatim}
  /* file: parser.ann */
static_consult([]).
residual(nont(_,_,_)).
filter(nont(X,T,R),[static,dynamic,dynamic]).
ann_clause(1,nont(X,T,R), (unfold(t(a,T,V)),memo(nont(X,V,R)))).
ann_clause(2,nont(X,T,R), (unfold(t(X,T,R)))).
ann_clause(3,t(X,[X|Es],Es),true).
\end{verbatim}
}

This supplies cogen with all the necessary information about the parser
program, this is, the code of the program (with annotations) and the result
of the binding-time analysis. The predicate \svbt{filter} defines the
division for the program and the predicate \svbt{residual} represents the
set ${\cal L}$ in the following way. If \svbt{residual(}$A$\svbt{)}
succeeds for a call $A$ then the predicate symbol $p$ of $A$ is in
$\mathit{Pred}(P)\backslash {\cal L}$ and $p$ is therefore one of the predicates for
which a $m$-predicate is going to be generated. The annotations
\svbt{unfold} and \svbt{memo} is used by cogen to determine whether or not
to unfold a call.

The generating extension produced by $\mathit{cogen}$ for the annotation
$nont(s,d,d)$ is:

{\footnotesize
\begin{verbatim}
/* file: parser.gx */
/*  --------------------  */
/*  GENERATING EXTENSION  */
/*  --------------------  */
:- logen_reconsult('memo').
:- logen_reconsult('pp').
nont_m(B,C,D,E) :- 
  (( find_pattern(nont(B,C,D),E)
   ) -> ( true ) ; (
    insert_pattern(nont(B,F,G),H),
    findall(I, (nont_u(B,F,G,J),I = (clause(H,J))),K),
    pp(K), find_pattern(nont(B,C,D),E)
  )).
nont_u(B,C,D,','(E,F)) :- t_u(a,C,G,E), nont_m(B,G,D,F).
nont_u(H,I,J,K) :- t_u(H,I,J,K).
t_u(L,[L|M],M,true).
\end{verbatim}
}

\noindent The generating extension is usually executed
 using {\tt\small nont\_m}, whereby the last argument is instantated
 to the filtered version of the call under consideration.
E.g., to specialise the original program for
{\tt\small nont(c,T,R)} we call
 {\tt\small nont\_m(c,T,R,FCall)}, which
 instantiates {\tt\small FCall} to
 {\tt\small nont\_\_0(T,R)} and prints
 the following residual program:

{\footnotesize
  \begin{verbatim}
  nont__0([a|B],C) :- 
    nont__0(B,C).
  nont__0([c|D],D).
\end{verbatim}
}

Observe that we can use the computed answer substitution
 for {\tt\small FCall} to produce an interface definition clause:
{\footnotesize
  \begin{verbatim}
  nont(c,T,R) :- nont__0(T,R).
\end{verbatim}
}
This will be done automatically by the {\sc logen} system
 when it produces the specialised program.

Some other examples which can be handled by simple divisions
(i.e., using just the binding-types {\tt\small static} and {\tt\small dynamic}),
 such as 
 an interpreter for the ground representation
 (where the overhead is compiled away)
 and a ``special'' regular expression parser
 from \cite{MogensenBondorf:LOPSTR92}
 (where we obtain deterministic automaton after specialisation)
 can be found in \cite{JorgensenLeuschel:Cogen}.


\section{The Transpose Example}\label{appendix:transpose}

A possible annotated program of the transpose benchmark program for matrix
 transposition looks like:

{\footnotesize
\begin{verbatim}
  static_consult([]).
  residual(transpose(A,B)).
  filter(transpose(A,B),[type(list(type(list(dynamic)))),dynamic]).
  ann_clause(1,transpose(A,[]),unfold(nullrows(A))).
  ann_clause(2,transpose(A,[B|C]),
                  (unfold(makerow(A,B,D)),unfold(transpose(D,C)))).
  filter(makerow(A,B,C),[type(list(type(list(dynamic)))),dynamic,dynamic]).
  ann_clause(3,makerow([],[],[]),true).
  ann_clause(4,makerow([[A|B]|C],[A|D],[B|E]),unfold(makerow(C,D,E))).
  filter(nullrows(A),[type(list(type(list(dynamic))))]).
  ann_clause(5,nullrows([]),true).
  ann_clause(6,nullrows([[]|A]),unfold(nullrows(A))).
\end{verbatim}}

In the above we stipulate that the first argument to transpose
 will be of type $list(list(\mathit{dynamic}))$, i.e., a list skeleton whose
  elements are in turn list skeletons (in other words we have a matrix 
  skeleton, without the actual matrix elements).
The generating extension produced by $\mathit{cogen}$ 
then looks like this:

{\footnotesize
\begin{verbatim}
/* file: bench/transpose.gx */
/*  --------------------  */
/*  GENERATING EXTENSION  */
/*  --------------------  */
:- logen_reconsult('memo').
:- logen_reconsult('pp').
transpose_m(B,C,D) :- 
  (( find_pattern(transpose(B,C),D)
   ) -> ( true ) ; (
    generalise(transpose(B,C),E), add_extra_argument([95,117],E,F,G),
    insert_pattern(E,H), findall(I, (G,I = (clause(H,F))),J),
    pp(J), find_pattern(transpose(B,C),D)
  )).
transpose_u(B,[],C) :-  nullrows_u(B,C).
transpose_u(D,[E|F],','(G,H)) :- makerow_u(D,E,I,G), transpose_u(I,F,H).
makerow_u([],[],[],true).
makerow_u([[J|K]|L],[J|M],[K|N],O) :-  makerow_u(L,M,N,O).
nullrows_u([],true).
nullrows_u([[]|P],Q) :- nullrows_u(P,Q).

\end{verbatim}}

\noindent Running the generating extension
 for {\tt\small transpose([[a,b],[c,d]],R)}
 leads to the following specialised
 program (and full unfolding has been achieved):

{\footnotesize
\begin{verbatim}
transpose([[a,b],[c,d]],A) :- transpose__0(a,b,c,d,A).
transpose__0(B,C,D,E,[[B,D],[C,E]]).
\end{verbatim}}

For the particular {\sc dppd} benchmark 
 query used in Section~\ref{section:benchmarks}
 we actually had to use a sligthly more refined division:

{\footnotesize
\begin{verbatim}
filter(transpose(A,B),  [ ( struct('[]',[])  ; 
     struct('.',[type(list(dynamic)),type(list(dynamic))])),dynamic]).
filter(makerow(A,B,C),[type(list(type(list(dynamic)))),dynamic,dynamic]).
\end{verbatim}}

The above corresponds to giving the first argument of {\tt\small transpose} the
 following binding-type (i.e., a list skeleton where only the first 
 argument itself is also a list skeleton):

{\footnotesize
\begin{verbatim}
:- type arg1 --> [] ; [list(dynamic) | list(dynamic)].
\end{verbatim}}

Using this division,
 the specialised program for {\tt\small transpose([[a,b],[c,d]],R)}
 is:
{\footnotesize
\begin{verbatim}
transpose([[a,b],[c,d]],A) :- transpose__0(a,b,[c,d],A).
transpose__0(B,C,[D,E],[[B,D],[C,E]]).
\end{verbatim}}

\end{document}